\documentclass[aps,prb,twocolumn,superscriptaddress]{revtex4-2}
\usepackage{bm}
\usepackage{graphicx}
\usepackage{mathrsfs}
\usepackage{amsmath,amssymb}
\usepackage{braket}
\usepackage{multirow}

\begin{document}
\title{Slowly generated anomalously large nuclear field in bulk $n$--AlGaAs}

\author{A.\ Shen}
\affiliation{Graduate School of Engineering, Hokkaido University, N13 W8, Kitaku, Sapporo 060-8628, Japan}
\author{J.\ Chen}
\affiliation{Graduate School of Engineering, Hokkaido University, N13 W8, Kitaku, Sapporo 060-8628, Japan}
\author{R.\ Kaji}
\email{r-kaji@eng.hokudai.ac.jp}
\affiliation{Graduate School of Engineering, Hokkaido University, N13 W8, Kitaku, Sapporo 060-8628, Japan}
\author{S.\ Yamamoto}
\affiliation{Department of Materials Science, Tohoku University, 6-6-02, Aoba-ku, Sendai 980-8579, Japan}
\author{H.\ Sasakura}
\affiliation{Graduate School of Engineering, Hokkaido University, N13 W8, Kitaku, Sapporo 060-8628, Japan}
\author{T.\ Uemura}
\affiliation{Graduate School of Information Science and Technology, Hokkaido University, N13 W8, Kitaku, Sapporo 060-8628, Japan}
\author{S.\ Adachi}
\email{adachi-s@eng.hokudai.ac.jp}
\affiliation{Graduate School of Engineering, Hokkaido University, N13 W8, Kitaku, Sapporo 060-8628, Japan}

\date{\today}

\begin{abstract}
	This study investigated the formation and relaxation dynamics of nuclear spin polarization in three Al$_x$Ga$_{1-x}$As bulk samples with different aluminum concentrations $x$ of 0.00, 0.05, and 0.15. The time-resolved Kerr rotation technique was primarily used. The samples with $x=$ 0.15 and 0.05 exhibited anomalously large nuclear magnetic fields $B_{\rm N}$ exceeding 1 T, approximately twice the applied magnetic field.
	Further investigations revealed that $B_{\rm N}$ formation  occurred in two-stages, a rapid initial rise followed by a gradual increase toward a saturation value. Relaxation measurements revealed that the relaxation time of $B_{\rm N}$ was longer for AlGaAs than for GaAs. The comparison of the results obtained under strong and weak magnetic fields indicated the suppression of quadrupole-induced relaxation. We modified the dynamics model of nuclear spin polarization and explained the two-stage formation and the accompanying large $B_{\rm N}$ in AlGaAs bulks.
\end{abstract}

\maketitle

\section{Introduction}

Considerable efforts have been devoted to the realization of spintronics and quantum information processing using nuclear spins~\cite{awschalom2002semiconductor}. In quantum processing, the nuclear spin can be utilized as a quantum bit (qubit) to store and modify quantum information~\cite{xu2023two,uysal2023coherent}. The nuclear spin is appealing for quantum processing applications as it is less vulnerable to perturbations and typically exhibits a longer coherence time than the electron spin. Consequently, research on reading the nuclear spin qubit with high fidelity~\cite{pla2013high}, transferring a quantum state from electron spin to nuclear spin ensemble~\cite{gangloff2019quantum,bodey2019optical,gangloff2021witnessing}, and the quantum sensing of a single nuclear-spin excitation (a nuclear magnon)~\cite{jackson2021quantum} continues to progress.

Optical pumping of nuclear spin polarization (NSP) in quantum wells (QWs) and quantum dots (QDs) has been studied extensively so far. In those nanostructured semiconductors, the hyperfine interaction (HFI), which couples electron and lattice nuclear spins, is enhanced due to the electron localization. Thus, a considerable NSP can be generated through spin transfer from photo-injected electrons to the lattice nuclei, providing the nuclear field that is an effective magnetic field for the localized electron spins.
A large nuclear field of $\sim 0.9$ T, comparable to an applied magnetic field of 1 T, has been reported in undoped GaAs/Al$_{0.35}$Ga$_{0.65}$As QWs~\cite{mocek2017high}. In the individual In(Ga)As and InAlAs QDs, where the electrons are fully localized, larger nuclear fields enough to cancel stronger applied field of $3-5$ T have been observed~\cite{urbaszek2013nuclear, PhysRevB.101.245424}.

Recently, the study of nuclear quadrupole interaction (NQI) has attracted much attention~\cite{huang2020oblique}. The NQI, resulting from the coupling of the electric field gradient (EFG) in the crystal and the electric quadrupole moment of nuclei with spin $I \ge 1$, modifies the energy structure of nuclear spins, leading to both the enhancement and suppression of NSP relaxation in complex ways. Especially in self-assembled QDs, the impacts of NQI are highly enhanced due to the residual strain. Thus, novel phenomena related to the NQI have been observed, such as the anomalous Hanle effects~\cite{Kreb_PRL2010, Yamamoto_PRB18, Yamamoto_PRB23} and the dragging effects of QD resonances~\cite{Latta_NatPhys2009, Hogele_PRL2012}.

Nanostructured semiconductors are advanced materials and a promising platform for applications in quantum information technology. However, the complexity of the structure and the associated inhomogeneity make it difficult to obtain quantitative agreement between experimental and theoretical results. 
Conversely, by using bulk semiconductors which have simple structures and excellent uniformity, it is possible to avoid this kind of problem and study in detail the effects of various interactions within the system (including NQI) on the dynamics of NSP.

The NSP in bulk GaAs and AlGaAs has been studied for several decades, where the Hanle effect measurements, the depolarization curve of time-integrated photoluminescence (PL) signals under a transverse magnetic field, have been widely used as a powerful tool for the study of electron-nuclei coupled system. Partial substitution of gallium atoms with aluminum atoms induces EFGs through symmetry lowering of nucleus sites so that nuclear quadrupole effects appear even in unstrained bulk AlGaAs~\cite{ihm1980valence}. In fact, clear nuclear quadrupole splittings at several hundred kHz have been observed via optically detected nuclear magnetic resonance of $p$-doped Al$_{0.26}$Ga$_{0.74}$As bulks~\cite{zakharchenya1981optical, OptOrientation84}, and interesting phenomena related to NQI, including the magnetic anisotropy and auto-oscillation of NSP, have been investigated so far~\cite{artemova1991sinergetic}.

More recently, a measurement technique based on electron spin noise spectroscopy has also been successful in the study of nuclear spin ensembles in semiconductors~\cite{SNS1,SNS2,SNS3}.
Spin noise spectroscopy (SNS), a detection scheme that does not involve the generation of nonequilibrium electron spin polarization (ESP), enables to extract only the relaxation process of NSP.
The presence of nonequilibrium ESP prevents complete separation of NSP formation and relaxation processes, which is an unavoidable problem with conventional techniques including Hanle effect measurements.

Previous studies in bulk samples were often performed under relatively weak magnetic fields of 0.1 T or less. In particular, in the measurements with SNS, low external fields of less than a few 10 mT have been widely used to evaluate the local magnetic field, which determines the heat capacity of nuclear spin systems~\cite{vladimirova2017nuclear,vladimirova2018spintemp}. However, there are few reports on how and to what extent NSPs form in bulk crystals under the influence of NQI in strong external magnetic fields (exceeding local magnetic fields).

In this paper, we investigate the formation and relaxation dynamics of NSP in bulk samples. Time-resolved Kerr rotation (TRKR) spectroscopy is used as a sensitive probe of nuclear spin system. In bulk AlGaAs samples, where NQIs are active, a large nuclear field above the external magnetic field and slow formation and relaxation dynamics are observed.

This paper is organized as follows: Section II provides the details of the samples and setup for the measurements. Section III describes the experimental results. 
After determining the electron g-factors, we show that the large nuclear fields appear only in the AlGaAs samples.
By modifying the pulse sequence of conventional TRKR measurements, we study the formation and relaxation dynamics of NSP in detail. 
Section IV presents a model that reproduces the observed results and discusses the experimental results compared to the model calculations. Section V is the conclusion of this paper.

\section{Samples and Experimental setup}

We used three $n$-Al$_x$Ga$_{1-x}$As bulk samples; all were grown on (001) GaAs substrates by molecular-beam epitaxy. The samples A and B had 1-$\mu$m-thick Si-doped $n$-AlGaAs layers with $x=0.15$ and $x=0.05$, respectively. Sample C had a Si-doped $n$-GaAs epitaxial layer ($x=0.00$) of thickness 472 nm. The AlGaAs samples A and B were commercial ones fabricated by QD Laser, Inc.~\cite{QDlaser}. The dopant density $n_{\rm d}$ of the three samples was $\sim 1 \times 10^{16}$ {\rm cm}$^{-3}$, which is just below the metal-insulator-transition region~\cite{vladimirova2017nuclear}. As shown later, the long coherence times of the resident electron spin polarization (RESP) in these samples make them suitable for the investigation of electron and nuclear spin dynamics via TRKR.

\begin{figure}
    \includegraphics[width=240 pt]{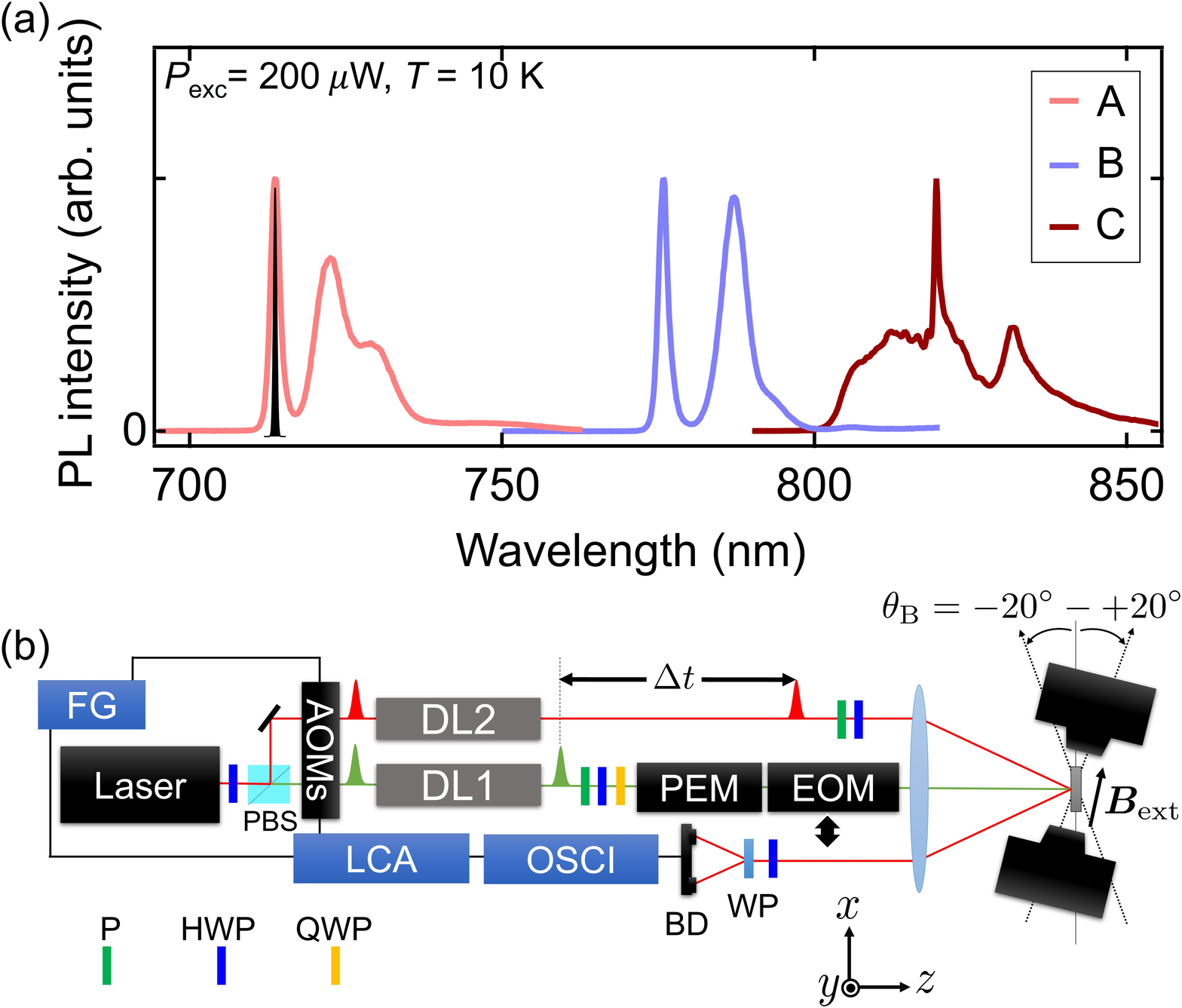}
    \caption{(a) Normalized PL spectra of samples A, B, and C obtained at 10 K by He--Ne laser excitation of $P_{\rm exc}=$ 200 $\mu$W. The black spectrum shows the excitation laser spectrum used in TRKR measurement for sample A. (b) Schematics of the experimental setup. See main text for abbreviations.}
    \label{Fig1}
\end{figure}

Prior to performing the TRKR experiments, all samples were subjected to PL measurements with a He--Ne laser (632.8 nm) to investigate the excitation power and temperature dependencies. Figure~\ref{Fig1}(a) shows the normalized PL spectra of the three samples measured at 10 K with an excitation power $P_{\rm exc}$ of 200 $\mu$W. The strongest sharp peak for each sample was located at 713.7 nm (A), 775.9 nm (B), and 819.5 nm (C), which belong to the emission range near the absorption band edge.

In the $P_{\rm exc}$-dependence, the PL intensity of the sharp peak increased linearly with $P_{\rm exc}$. In contrast, the broader peaks in the longer wavelength region tended to saturate for each sample. Furthermore, the sharp peak remained even at higher temperatures, whereas the broader peaks demonstrated faster relaxation. The PL energy shift was consistent with the temperature shift of the bandgap energy of Al$_x$Ga$_{1-x}$As of $x=0.15$ (A), $0.05$ (B), and $0.00$ (C), respectively, in the observed region (4.2 to 100 K)~\cite{vurgaftman2001band}. Since Si dopants are known to produce shallow donors in (Al)GaAs~\cite{vladimirova2017nuclear}, the sharp and broad PL peaks are attributed to donor-bound excitons and donor-acceptor recombination lines~\cite{PhysRevB.66.245204}, respectively.

To observe the electron and nuclear spin dynamics directly, we utilized the TRKR  method. Figure~\ref{Fig1}(b) shows a schematic of the TRKR setup. The mode-locked Ti:sapphire laser with the pulse repetition period of $\sim$13.1 ns (Coherent Inc., MIRA900D) provided the optical excitation and probing of the studied samples.
The wavelength of the laser was adjusted to the sharp peaks of each sample.
The laser spectrum for sample A, with full width at half maximum of 0.44 nm, is also presented as an example in Fig. 1(a). The spectral width corresponded to a pulse width of $\sim$2 ps in the Fourier limit, which was consistent with the autocorrelation measurement of the laser pulse.

As in the standard one-color TRKR (or TR-Faraday rotation) system~\cite{baumberg1994ultrafast,crooker1997optical, Kikkawa1997,kennedy2006optical}, the laser beam was divided into two using a polarizing beam splitter (PBS) after passing through a half-wave plate (HWP) that adjusted the power ratio of the pump and probe beams. A circularly-polarized ($\sigma^{+}$ or $\sigma^{-}$) pump pulse was incident along the sample growth axis ($z$ axis). An external magnetic field $\left| \bm{B}_{\rm ext} \right|$ up to 560 mT was applied within the angular range $\theta_{\rm B}$ of $\sim \pm 20^{\circ}$ from the $x$ axis. 
Linearly polarized probe pulses (with delay time $\Delta t$) monitor the time evolution of ESP excited by preceding pump pulses in the spot area ($\sim 150$ $\mu$m diameter).
$\Delta t$ was controlled by adjusting the length of delay lines. The probe delay line (DL2) allowed a maximum $\Delta t$ of 6.6 ns. Further, a total observation length of 12.6 ns was also achieved by changing the pump delay line (DL1).

A Kerr rotation angle on the probe beam was measured by a polarization bridge and lock-in detection technique. The polarization bridge comprised a HWP, a Wollaston prism (WP), and a balanced photodetector (BD) connected to an oscilloscope (OSCI). Here, the oscilloscope served as an amplifier and signal monitor. To improve the signal-to-noise ratio, the pump and probe beams were modulated by acousto-optic modulators (AOMs, Isomet, 1205C-1) with modulation frequencies of 250 and 130 kHz, respectively, which were driven by a 2-channel function generator (FG, NF Corporation, WF1968).

A lens ($f {=}$300 mm) focused the pump and probe beams on the sample surface mounted in the closed-cycle cryostat. The pump beam was incident normal to the sample surface, and the intersection angle between the pump and probe was less than $2^{\circ}$. The sample temperature was maintained at 10 K during all TRKR experiments. 
In Section~\ref{sec3-B}, to observe the formation and relaxation dynamics of NSP and the resultant nuclear field $\bm{B}_{\rm N}$, \textit{pre-pump-time} and \textit{dark-time} are introduced in the pulse sequence before scanning the delay time.

To clarify the difference depending on the presence and absence of NSP, TRKR measurements were performed in two patterns: $\sigma^{+}$ or $\sigma^{-}$ excitation by applying a constant voltage to an electro-optic modulator (EOM, ThorLabs, EO-AM-NR-C1), and 50-kHz polarization modulation with a photoelastic modulator (PEM, Hinds Instruments, PEM100). 
When using EOM, the difference in modulation frequency between pump and probe beams, 120 kHz,  was used as the reference signal for the lock-in amplifier (LCA, Signal Recovery, model 7265). When using PEM, the EOM was shifted out of the pump path, and a reference frequency of 30 kHz, that is, the difference between the probe modulation and twice the PEM operating frequency, was used. A pair of quarter-wave plate (QWP) and a HWP were placed in front of the PEM to adjust the input polarization and to correct phase distortion induced by the optical elements.

\section{EXPERIMENTAL RESULTS}

\subsection{TRKR signals and electron g-factor in $\bm n$-AlGaAs}\label{sec3-A}

\begin{figure}[tb]
    \centering
    \includegraphics[width=240 pt]{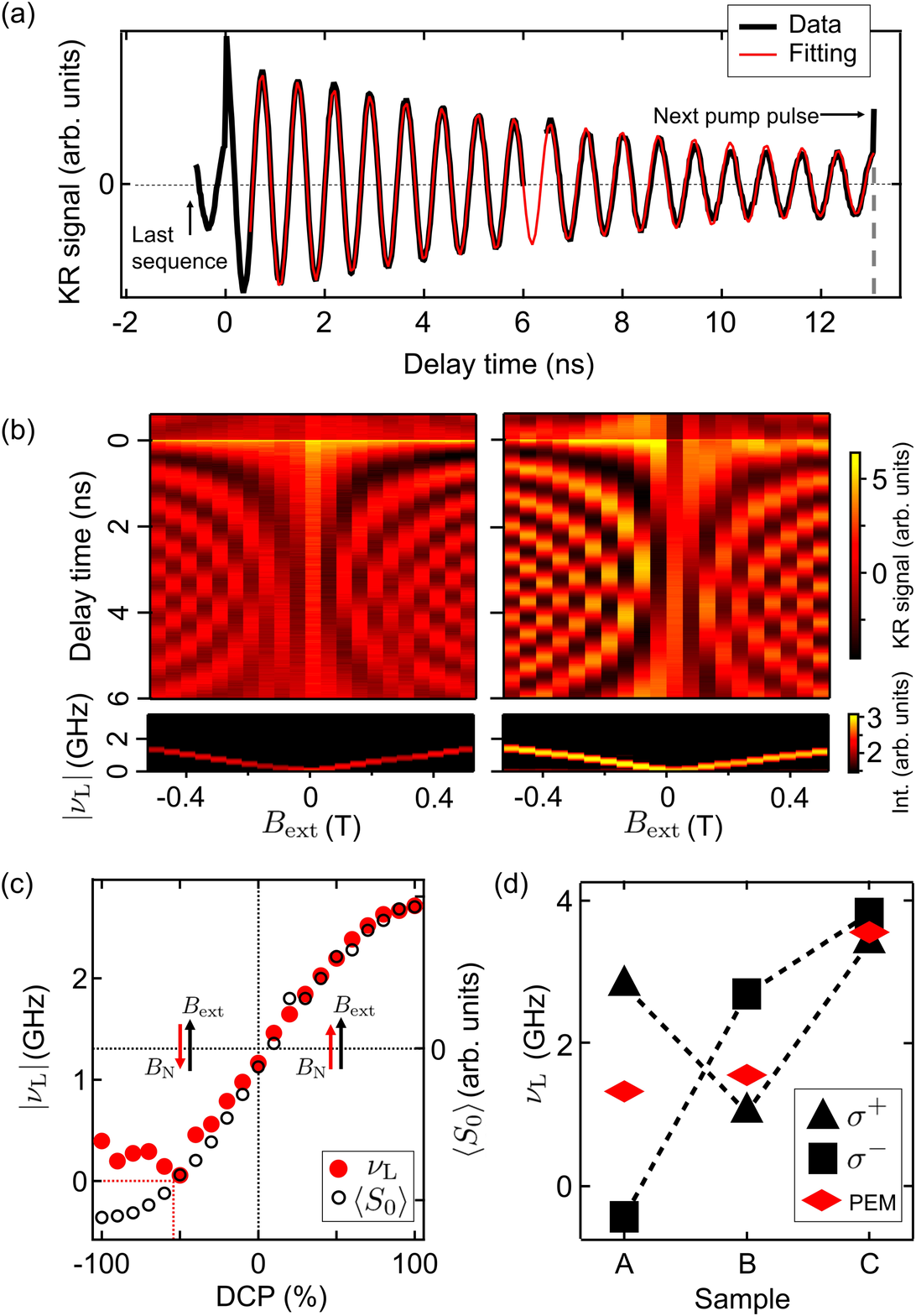}
    \caption{Example data in sample A. (a) The TRKR signal under following conditions: pump/probe power $= 10/1$ mW, temperature = 10 K, $B_{\rm ext}=+560$ mT, and $\theta_{\rm B}=0^{\circ}$. (b) Dependence of TRKR signals on $|B_{\rm ext}| \leq 560 \ {\rm mT}$ at $\theta_{\rm B} = 0^{\circ}$ (left) and 18$^{\circ}$ (right) with PEM. The bottom panels show the corresponding Fourier spectra.
    (c) The observed $|\nu_{\rm L}|$ (solid circles) and $\braket{S_0}$ (open circles) as a function of the pump DCP at $B_{\rm ext}=+560$ mT and $\theta_{\rm B}=+20^{\circ}$. The DCP of $+100\% \ (-100\%)$ corresponds to $\sigma^+ \ (\sigma^-)$ excitation. Inset depicts the direction of $B_{\rm N}$ relative to $B_{\rm ext}$. (d) Comparison of $\nu_{\rm L}$ under $\sigma^+$ (triangles) and $\sigma^-$ (squares) excitations in samples A, B, and C. $B_{\rm ext}$ and $\theta_{\rm B}$ are the same as in (c). 
    Red diamonds are the $\nu_{\rm L}$ obtained using PEM so that $B_{\rm N} =0$.
    }
    \label{Fig2}
\end{figure}

In Voigt geometry ($\theta_{\rm B}=0^{\circ}$), the electron spins injected parallel to the $z$ axis begin to precess with the Larmor frequency $\nu_{\rm L}$ in the $yz$-plane immediately after the optical excitation. An example of the observed TRKR signal in sample A at 10 K and its fitting are presented in Fig.~\ref{Fig2}(a).
The first part of the example data was within the range $-0.6 \sim 6$ ns, whereas the second part was from $- 6.6$ (6.5) ns to 0 (13.1) ns. A TRKR signal generally contains multi-spin components originating from neutral and charged excitons and RESP. The radiative lifetime of the excitons and their complexes limits the relaxation time within 1 ns~\cite{colton2002spin}, whereas the RESP has a long coherence time~\cite{SpinPhysicsinSC, PhysRevLett.80.4313}. In Fig.~\ref{Fig2}(a), the second part of the TRKR signal was connected to the one from the next pump pulse, implying that the RESP in this sample lasted longer than the pulse repetition period.

We used a fitting function to extract electron spin information from TRKR signals:
\begin{align}
\braket{S_z(\Delta t)} = \braket{S_0}{\rm exp} \left( -\frac{\Delta t}{\tau_{\rm d}} \right){\rm cos} \left( 2\pi\nu_{\rm L} \Delta t + \phi \right) + C,
\end{align}
where $\langle S_{z} \rangle$ is the projection of ESP onto the probe beam (almost the same as the $z$ axis). $\braket{S_0}$ is the initial ESP by the pump pulse, $\tau_{\rm d}$ is the decay time of the envelope of the oscillating TRKR signal and corresponds to  the electron spin lifetime, $T_2^*$, under normal conditions, $\phi$ is the initial phase, and $C$ is the zero signal level of the measurement.
As shown by the red line, this equation fits both parts of the data well, yielding $\tau_{\rm d}$ = 9.6 $\pm$ 0.2 ns.
Note that if $\tau_{\rm d}$ is comparable to or longer than the pulse repetition period, the $\tau_{\rm d}$ does not reflect 
the $T_{2}^{*}$ and a special method such as resonant spin amplification~\cite{PhysRevLett.80.4313, PhysRevB.85.121303} is required to evaluate the exact value of $T_{2}^{*}$.  Since the exact value of $T_{2}^{*}$ is not necessary for the discussion in this paper, we focus only on $\nu_{\rm L}$, which serves as a measure of $|\bm B_{\rm N}|$.

In the TRKR measurements, the observed $\nu_{\rm L}$ is determined by the following relation,
\begin{align}
	\nu_{\rm L} = |{\rm g}_{\rm e} \mu_{\rm B} (\bm{B}_{\rm ext} + \bm{B}_{\rm N})| /h,
    \label{eq1}
\end{align}
where ${\rm g}_{\rm e}$ is the electron g-factor, $\mu_{\rm B}$ is the Bohr magneton, and $h$ is the Planck constant.
Under the condition where $\bm{B}_{\rm N}$ is generated, the presence of $\bm{B}_{\rm N}$ is reflected in the change in $\nu_{\rm L}$.
It should be noted that TRKR measurements primarily yield the absolute value of $\nu_{\rm L}$. So, careful consideration is required when determining the signs of ${\rm g}_{\rm e}$ and $\left( \bm{B}_{\rm ext} + \bm{B}_{\rm N} \right)$.

First, the electron g-factor measurement was conducted, which is essential for evaluating $\bm{B}_{\rm N}$. In order to determine the \textit{intrinsic} g-factor, it is necessary to suppress $\bm{B}_{\rm N}$ formation completely. Thus, we modulated the pump helicity between $\sigma^{+}$ and $\sigma^{-}$ at the frequency of 50 kHz by using a PEM. 
Figure~\ref{Fig2}(b) presents two-dimensional plots of the TRKR signals of sample A as a function of $|\bm B_{\rm ext}|$ up to $\pm 560$ mT at $\theta_{\rm B}=0^{\circ}$ (left) and $18^{\circ}$ (right).
The bottom panels are the Fourier spectra of the TRKR signals in the upper panels.

Under the oblique $\bm B_{\rm ext}$, $\bm{B}_{\rm N}$ is generated by the non-precessing component of ESP $\langle \bm S_{\parallel} \rangle$ via the HFI. 
However, the TRKR signals and the correponding Fourier patterns differed minimally between the two $\theta_{\rm B}$. It indicates that the modulation period of 20 $\mu$s was short enough to suppress $\bm{B}_{\rm N}$ formation. It also suggests that the electron ${\rm g}$-factor was nearly isotropic. Details regarding $\bm{B}_{\rm N}$ formation under an oblique $\bm B_{\rm ext}$ and g-factor isotropy are presented in Appendix~\ref{Append_A}.
Under this condition, we confirmed that $\nu_{\rm L}$ was proportional to $\left| B_{\rm ext} \right|$, and found that $\left| {\rm g}_{\rm e} \right|$ for sample A was $0.170 \pm 0.001$ from a line fitting.
Similarly, $|{\rm g}_{\rm e}|$ for samples B and C were evaluated as 0.205 $\pm$ 0.001 and 0.447 $\pm$ 0.001, respectively.

Next we determine the sign of ${\rm g}_{\rm e}$ by utilizing $\bm{B}_{\rm N}$.
The constant pump helicity realized by an EOM polarizes nuclear spins, and the resultant $\bm{B}_{\rm N}$ causes the change in $\nu_{\rm L}$.
Figure~\ref{Fig2}(c) shows the pump helicity dependence of $\left| \nu_{\rm L} \right|$ in sample A.
By changing the applied voltage on the EOM, the degree of circular polarization (DCP) of the pump beam was continuously varied from $-100\%$ ($\sigma^-$) to $+100\%$ ($\sigma^+$).
The change in the pump DCP was reflected in $\langle S_{0} \rangle$, which exhibited the maximum (minimum) value at $+100\%$ ($-100\%$).
It was found that $\left| \nu_{\rm L} \right|$ showed a maximum value at $+100\%$ and decreased monotonically as the pump DCP decreased.
$\left| \nu_{\rm L} \right|$ reached zero at $\sim -55\%$ and increased again when the pump DCP was reduced further.

The increase (decrease) in $\left| \nu_{\rm L} \right|$ observed with positive (negative) DCP indicates an increase (decrease) in $\left| \bm{B}_{\rm ext} + \bm{B}_{\rm N} \right|$, that is, $\bm{B}_{\rm N}$ is in the same (opposite) direction as $\bm{B}_{\rm ext}$. 
Here, $\bm{B}_{\rm N}$ is defined as $\bm{B}_{\rm N} = 2 \tilde{A} \langle \bm{I} \rangle / ({\rm g}_{\rm e} \mu_{\rm B})$, where $\tilde{A}$ is the averaged hyperfine constant and is positive for III-V semiconductors~\cite{chekhovich2013element}. As shown in Fig.~\ref{Fig1A}(a) in Appendix A, $\langle \bm{S}_{\parallel} \rangle$ appears in the same (opposite) direction as $\bm{B}_{\rm ext}$ under $\sigma^{+}$ ($\sigma^{-}$) excitation. Further, if non-collinear HFI is negligible, the direction of $\langle \bm{I} \rangle$ under $\sigma^{+}$ ($\sigma^{-}$) excitation is also the same (opposite) as $\bm{B}_{\rm ext}$ from the conservation law in spin transfer.
Therefore, for $\bm{B}_{\rm N}$ to be opposite to $\bm{B}_{\rm ext}$ under negative pump DCP, ${\rm g}_{\rm e}$ must be positive in sample A.

The fact that $\left| \nu_{\rm L} \right|$ began to increase again at DCP $\lesssim -55\%$ suggests that $\bm{B}_{\rm N}$ overcompensating for $\bm{B}_{\rm ext}$ was generated. Details of formation and relaxation of such a large $|\bm{B}_{\rm N}|$ will be discussed in the next section.

Figure~\ref{Fig2}(d) summarizes the observed $\nu_{\rm L}$ in samples A, B, and C under $\sigma^+$ and $\sigma^-$ excitations. 
In contrast to sample A, $\nu_{\rm L}$ under $\sigma^+$ ($\sigma^-$) excitation is smaller (larger) than that obtained with the PEM in sample B and C.
This clearly shows that the signs of ${\rm g_e}$ for samples B and C are negative. 
Therefore, we conclude the values of electron g factor, including their signs, in samples A, B, and C as $+0.170\pm0.001$, $-0.205\pm0.001$, and $-0.447\pm0.001$, respectively.
These values are in good agreement with theoretical calculations in terms of both signs and magnitudes~\cite{weisbuch1977optical,chadi1976gamma,yugova2007universal}.

\subsection{Formation and relaxation dynamics of nuclear field}\label{sec3-B}

\begin{figure}[tb]
    \centering
    \includegraphics[width=240 pt]{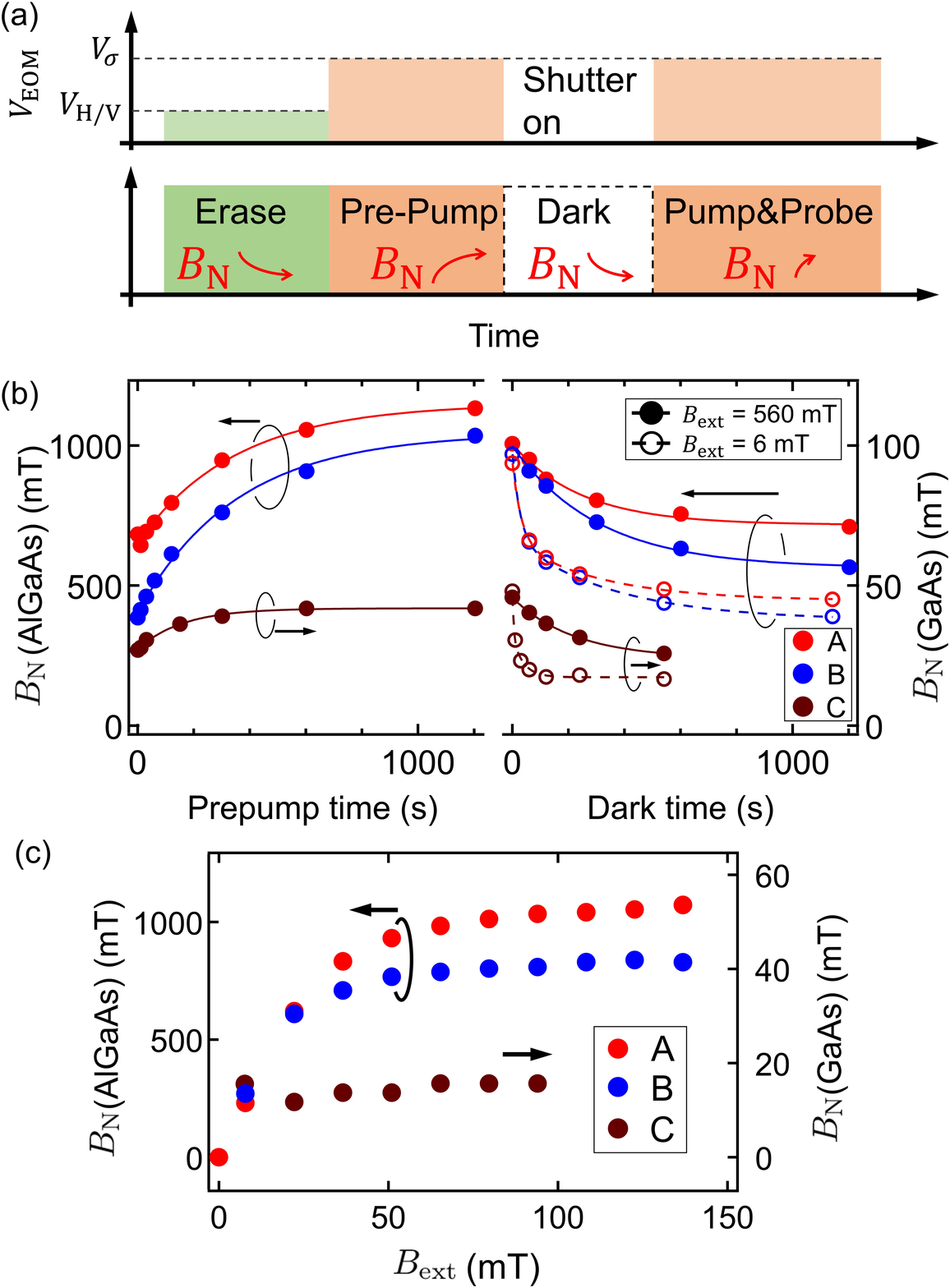}
    \caption{(a) Employed polarization sequence (bottom panel) and corresponding applied voltage to EOM (upper panel). (b) Formation (left) and relaxation (right) dynamics of ${B}_{\rm N}$ of the three samples at $\theta_{\rm B} =+20^{\circ}$. Formation dynamics were measured only under $B_{\rm ext}=+560$ mT, while relaxation dynamics were under +560 mT (solid circles) and +6 mT (open circles).  The solid and dashed curves are the fitting results. (c) Dependence of $B_{\rm N}$ on $B_{\rm ext}$ at $\theta_{\rm B} = +20^{\circ}$ for all samples. In (b) and (c), the left axis is set for samples A and B (AlGaAs), and the right axis is for sample C (GaAs). }
    \label{Fig3}
\end{figure}

In this section, we examine the formation and relaxation dynamics of the observed large ${\bm B}_{\rm N}$.
First, we focus on the formation dynamics of NSP. For this purpose, \textit{erase} and \textit{pre-pump} pulses were inserted before the regular pump and probe pulses for TRKR measurements. The pulse sequences and accompanying changes in $B_{\rm N}$ are summarized in Fig.~\ref{Fig3}(a).

To eliminate the remaining NSP and set the initial $\langle \bm{I} \rangle$ to $\bm{0}$ for each measurement, the sample was illuminated with linearly polarized erase pulses for $T_{\rm erase}=600$ s. Thereafter, the pulse polarization was switched to circular polarization for pre-pump pulses, generating NSP continuously during the interval $T_{\rm prep}$. The formation dynamics of NSP was revealed through the change in $\nu_{\rm L}$ associated with varying $T_{\rm prep}$. The light polarizations of the erase, pre-pump, and pump pulses were controlled through the input voltage of the EOM.

\begin{table*}[ht]
    \centering
    \caption{Fitting results from the formation and relaxation of ${B}_{\rm N}$ in three samples.}
    \label{Table}
    \begin{tabular}{cccc||cccc|ccc}
        \hline
        Formation                                    & \multicolumn{3}{c||}{${B}_{\rm ext}$ = 560 mT} & Relaxation    & \multicolumn{3}{c|}{${B}_{\rm ext}$ = 560 mT} & \multicolumn{3}{c}{${B}_{\rm ext}$ = 6 mT}                                                                                            \\
        \hline
        Sample                                       & A                                              & B             & C                                             & Sample                                        & A            & B            & C          & A            & B            & C            \\
        \hline
        $B_{\rm N}^{\rm max}$ (mT)                   & $1150 \pm 53$                                  & $1089 \pm 32$ & $43 \pm 1$                                    &                                               &              &              &            &              &              &              \\
        $B_{\rm f}^{\rm Form}$ (mT)                  & 567 $\pm$ 40                                   & 450 $\pm$ 20  & 33 $\pm$ 1                                    & $B_{\rm N}^{\rm offset}$ (mT)                 & 718 $\pm$ 13 & 564 $\pm$11  & 24 $\pm$1  & 447 $\pm$ 8  & 377 $\pm$ 11 & 18 $\pm$ 1   \\
        $\tau_{\rm f}^{\rm Form}$ (s)                & $-$                                            & $-$           & $-$                                           &                                               &              &              &            &              &              &              \\
                                                     &                                                &               &                                               & $B_{\rm f}^{\rm Relax}$ (mT)                  & $-$          & $-$          & $-$        & 279 $\pm$ 25 & 303 $\pm$ 23 & 14 $\pm$ 11  \\
                                                     &                                                &               &                                               & $\tau_{\rm f}^{\rm Relax}$ (s)                & $-$          & $-$          & $-$        & 32 $\pm$ 6   & 28 $\pm$ 5   & 5 $\pm$ 7    \\
        $B_{\rm s}^{\rm Form}$ (mT)                  & 582 $\pm$ 35                                   & 638 $\pm$ 25  & 10 $\pm$ 1                                    & $B_{\rm s}^{\rm Relax}$ (mT)                  & 287 $\pm$ 16 & 435 $\pm$ 12 & 22 $\pm$ 1 & 210 $\pm$ 21 & 289 $\pm$ 18 & 14 $\pm$ 10  \\
        $\tau_{\rm s}^{\rm Form}$ (s)                & 327 $\pm$ 60                                   & 487 $\pm$ 60  & 130 $\pm$ 13                                  & $\tau_{\rm s}^{\rm Relax}$ (s)                & 248 $\pm$ 38 & 306 $\pm$ 24 & 221$\pm$9  & 316 $\pm$ 64 & 362 $\pm$ 61 & 38 $\pm$  31 \\
        \hline
    \end{tabular}
\end{table*}

$T_{\rm prep}$-dependence of $B_{\rm N}$ is summarized on the left side of Fig.~\ref{Fig3}(b).  The measurements were performed under $B_{\rm ext} = +560$ mT at $\theta_{\rm B} = +20^{\circ}$. In all samples, the observed values of $B_{\rm N}$ increased with increasing $T_{\rm prep}$ and reached their maximum values, $B_{\rm N}^{\rm max}$ (e.g., 1150 mT in sample A) within $T_{\rm prep}=1200$ s. It is noteworthy that the formation dynamics occur in two stages: a rapid initial rise and a subsequent gradual increase toward saturation value. In samples A and B ($x=0.15$ and 0.05), the initial rise of $B_{\rm N}$  ($\sim 570$ mT and $\sim 450$ mT, respectively) reached comparable to $B_{\rm ext}$. The initial rise of $B_{\rm N}$ was also observed in sample C ($x=0.00$), but the value was only $\sim 30$ mT.
After the initial rise, $B_{\rm N}$ increased gradually, but the rate of increase was different for AlGaAs (A and B) and GaAs (C). For samples A and B, it took more than 1000 s to reach the saturation value, while it took only $\sim 300$ s for sample C.
Similar two-stage NSP formation has been reported in undoped single GaAs/AlGaAs QW~\cite{malinowski2000nuclear,malinowski2001nuclear} under $B_{\rm ext} \ge$1 T.

We attempted to fit the experimental results with the following double exponential function:
\begin{align}
    B_{\rm N}(T_{\rm prep}) & =B_{\rm f}^{\rm Form} \left[1- \exp(-T_{\rm prep}/\tau_{\rm f}^{\rm Form}) \right]  \nonumber    \\
                            & + B_{\rm s}^{\rm Form} \left[1- \exp(-T_{\rm prep}/\tau_{\rm s}^{\rm Form}) \right]. \label{eq2}
\end{align}
Here, $B_{\rm f}^{\rm Form}$ ($B_{\rm s}^{\rm Form}$) and $\tau_{\rm f}^{\rm Form}$ ($\tau_{\rm s}^{\rm Form}$) are the amplitude and rise time constant of the fast (slow) component, respectively. Although this expression represents a two-stage $B_{\rm N}$ formation as a function of $T_{\rm prep}$, it is difficult to deduce $\tau_{\rm f}^{\rm Form}$ accurately for now. 
This is because $\tau_{\rm f}^{\rm Form}$ seems to be shorter than the time resolution of measurement, as non-zero $B_{\rm f}^{\rm Form}$ was observed at $T_{\rm prep}=0$. 
Hence, we set $\tau_{\rm f}^{\rm Form}=0$ and fitted the slow component by a single exponential function with an offset $B_{\rm f}^{\rm Form}$. The solid curves in Fig.~\ref{Fig3}(b) depict the fitting results. The obtained parameters ($B_{\rm s}^{\rm Form}$, $\tau_{\rm s}^{\rm Form}$, $B_{\rm f}^{\rm Form}$) are summarized in the left side of Table~\ref{Table}. $B_{\rm N}^{\rm max}$ in Table~\ref{Table} is the sum of $B_{\rm f}^{\rm Form}$ and $B_{\rm s}^{\rm Form}$.

The formation time $\tau_{\rm s}^{\rm Form}$ of the slow component is significantly longer in samples A and B than in sample C. Furthermore, there is a significant difference in the relative values of $B_{\rm f}^{\rm Form}$ and $B_{\rm s}^{\rm Form}$ between AlGaAs and GaAs. The ratios, $B_{\rm s}^{\rm Form}/B_{\rm f}^{\rm Form}$, are about 1.0 and 1.4 in samples A and B,  respectively. This indicates that the slow component contributes as much or more than the fast component to $B_{\rm N}^{\rm max}$ in the AlGaAs bulks.
In sample C, the ratio is only about 0.3, indicating that the fast component (or the initial rise) is dominant.
These differences strongly indicate that Al-incorporation plays an important role in the dynamics of NSP formation.

Now that $T_{\rm prep}$ for NSP saturation has been determined, the next step is to examine the effect of $B_{\rm ext}$.
Figure~\ref{Fig3}(c) shows the observed $B_{\rm N}$ as a function of $B_{\rm ext}$. For all samples, $T_{\rm prep}$ and $\theta_{\rm B}$ were fixed to 1200 s and $+20^{\circ}$ to ensure that $B_{\rm N}$ maintained its saturation value at each $B_{\rm ext}$. 
In samples A and B, $B_{\rm N}$ was found to increase with increasing $B_{\rm ext}$ in the region below 50 mT, but to remain constant above 50 mT. 
On the other hand, $B_{\rm N}$ in sample C was almost constant above 6 mT, the weakest $\left| B_{\rm ext} \right|$ at which $\nu_{\rm L}$ could be evaluated accurately. These experimental facts suggest that the nuclear spin relaxation rate in AlGaAs varies with $B_{\rm ext}$; it is related to the impact of NQI, as discussed later.

To confirm this scenario, the relaxation dynamics of ${\bm B}_{\rm N}$ was measured by adding \textit{dark-time} ($T_{\rm dark}$) after the pre-pump pulses as shown in Fig.~\ref{Fig3}(a). Under the conditions $T_{\rm erase}=600$ s, $T_{\rm prep}=1200$ s, $B_{\rm ext}=560$ mT, and $\theta_{\rm B} =+20^{\circ}$, we confirmed that almost identical ${B}_{\rm N}$ was generated in each measurement. Then, all laser beams were blocked during $T_{\rm dark}$, and changes in $B_{\rm N}$ depending on $T_{\rm dark}$ were examined. Two values of $B_{\rm ext}$ (560 mT and 6 mT) were employed as the one during $T_{\rm dark}$, because the NSP relaxation in AlGaAs was expected to vary whether $B_{\rm ext}$ was strong or weak.

The right side of Fig.~\ref{Fig3}(b) shows the $B_{\rm N}$ obtained for strong (560 mT) and weak $B_{\rm ext}$ (6 mT) as a function of $T_{\rm dark}$. For all data sets, the observed $B_{\rm N}$ became smaller as $T_{\rm dark}$ increased, but did not reach zero even in the long-$T_{\rm dark}$ limit (i.e., an offset $B_{\rm N}^{\rm offset}$ occurred). 
This is because the nuclear spins, which had been relaxed during $T_{\rm dark}$, were re-polarized by ESP during the TRKR measurement stage; that time scale seems to be comparable to the fast formation time $\tau_{\rm f}^{\rm Form}$.

The relaxation under $B_{\rm ext} =$ 560 mT (solid circles) was fitted well by a single exponential function with the time constant $\tau_{\rm s}^{\rm Relax}$.
In contrast, for the relaxation under $B_{\rm ext} =$ 6 mT (open circles), it was necessary to use the following double exponential function for fitting:
\begin{align}
    B_{\rm N}(T_{\rm dark}) & = B_{\rm N}^{\rm offset} + B_{\rm f}^{\rm Relax} \exp \left( - \frac{T_{\rm dark}}{\tau_{\rm f}^{\rm Relax}} \right) \nonumber \\
                            & +  B_{\rm s}^{\rm Relax} \exp \left( -\frac{T_{\rm dark}}{\tau_{\rm s}^{\rm Relax}} \right), \label{eq3}
\end{align}
where the fast (slow) component with the amplitude $B_{\rm f}^{\rm Relax}$ ($B_{\rm s}^{\rm Relax}$) relaxed with the time constant $\tau_{\rm f}^{\rm Relax}$ ($\tau_{\rm s}^{\rm Relax}$).
The fitting results are summarized on the right part of Table~\ref{Table}.

\section{DISCUSSION}

\subsection{Model for NSP dynamics and calculated results}
In general, the temporal evolution of NSP, $\langle \bm I_{\parallel} \rangle$, is described by the following rate equation~\cite{OptOrientation84},

\begin{align}
    \frac{d\langle \bm I_{\parallel} \rangle}{dt} = \frac{1}{T_{\rm F}} \left[ Q\langle \bm S_{\parallel} \rangle - \langle \bm I_{\parallel} \rangle \right] - \frac{1}{T_{\rm R}}\langle \bm I_{\parallel} \rangle,
    \label{eq4}
\end{align}
where $Q = {\overline {I(I + 1)}}/[S(S + 1)]$ is the spin conversion coefficient between electron and nucleus. 
$1/T_{\rm F}$ represents the spin transfer rate via HFI, while $1/T_{\rm R}$ is the relaxation rate due to the dipole interaction among nuclei and NQI, as well as other interactions.

This simple equation has successfully explained various experimental results such as bistability~\cite{urbaszek2013nuclear} and tristability~\cite{PhysRevB.101.245424} of NSP observed in single QDs under longitudinal magnetic fields. This equation also suggests that NSP exhibits a single exponential time evolution if the rates $1/T_{\rm F}$ and $1/T_{\rm R}$ are regarded as constants, as reported in undoped GaAs/AlGaAs QWs with very low impurity density~\cite{mocek2017high}. However, the two-stage NSP formation, as shown in Fig.~\ref{Fig3}(b), cannot be reproduced.

\begin{figure}[tb]
    \centering
    \includegraphics[width=240pt]{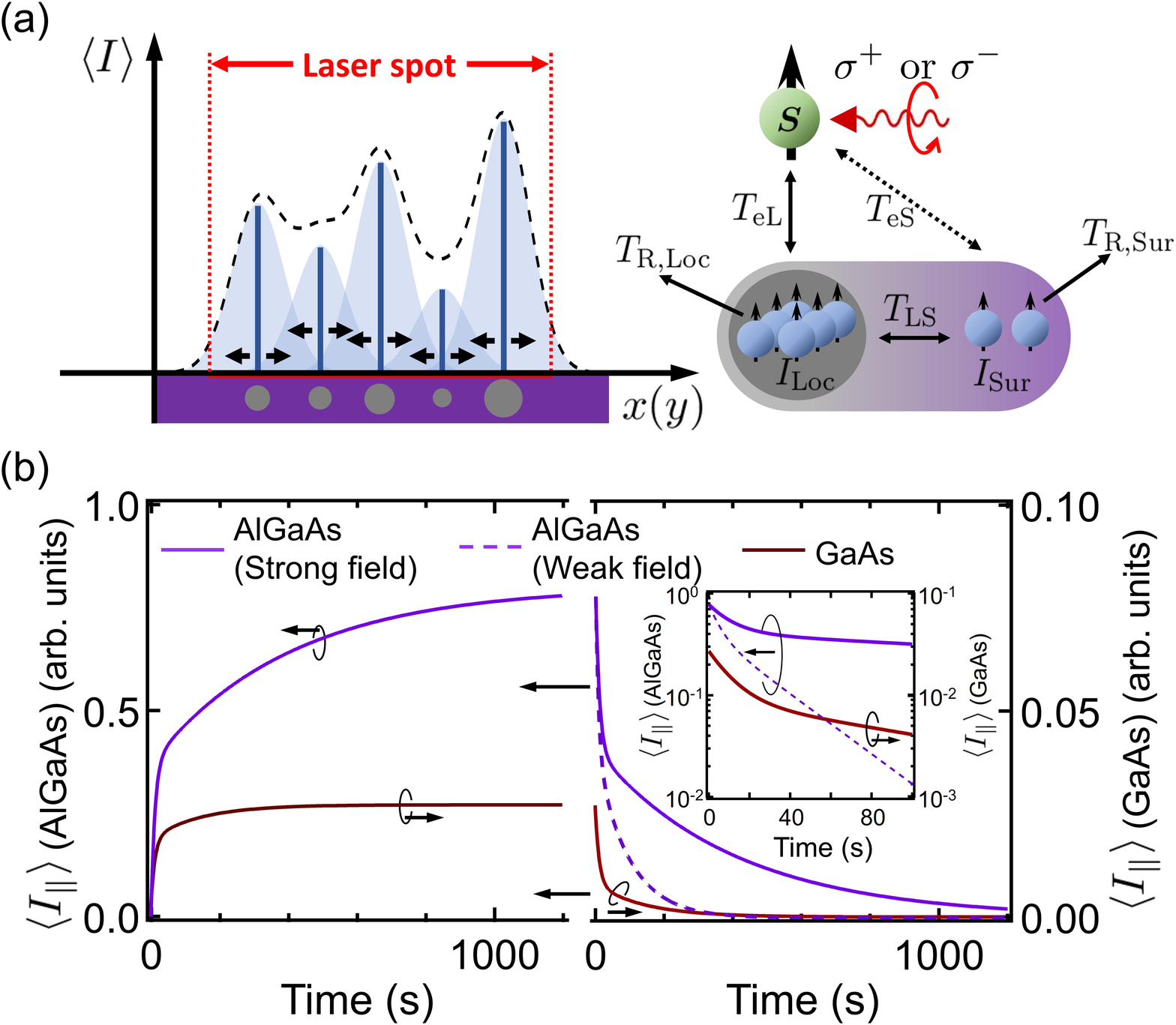}
    \caption{(a) Schematic diagram of NSP formation and spin transfer in the system (left) and the model under consideration (right). (b) Model calculations of the formation (left) and relaxation(right) processes of NSP. Solid (dashed) curves represent the results under $|B_{\rm ext}| \gg (<) B_{\rm L}$. In the calculation of the relaxation process, the fast re-polarized NSP that causes  an offset is ignored. 
    The inset shows the same calculations of relaxation dynamics on a logarithmic scale to demonstrate the two components more clearly.}
    \label{Fig4}
\end{figure}

In order to reproduce the observations, we divided the nuclear spin ensemble within the laser spot ($\sim$150-$\mu$m-diameter) into two groups, nuclear spins near the electron localization sites and those in the surrounding region, and considered the temporal evolution of each group. The former group constitutes the NSP labeled as $\langle \bm{I}_{\rm Loc} \rangle$, and the latter corresponds to $\langle \bm{I}_{\rm Sur} \rangle$. An overview of the model is shown in Fig.~\ref{Fig4}(a).
Under circularly polarized excitation, the spin transfer from photo-oriented electrons to nuclear spins occurs.
However, the strength of the HFI depends on proximity of the nuclear spins to the electron localization sites.
Therefore, the two groups have different spin transfer rates: $1/T_{\rm eL}$ for $\langle \bm{I}_{\rm Loc} \rangle$ and $1/T_{\rm eS}$ for $\langle \bm{I}_{\rm Sur} \rangle$. Given the HFI strength for each group, these spin transfer rates should be $1/T_{\rm eL} \gg 1/T_{\rm eS}$.
Then, $\langle \bm{I}_{\rm Loc} \rangle$ and $\langle \bm{I}_{\rm Sur} \rangle$ relax at relaxation rates $1/T_{\rm R, Loc}$ and $1/T_{\rm R, Sur}$, respectively, and spin transfer occurs between them, resulting in mutual time evolution.
We labeled the spin transfer rate between $\langle \bm{I}_{\rm Loc} \rangle$ and $\langle \bm{I}_{\rm Sur} \rangle$ as $1/T_{\rm LS}$.

Consequently, the rate equations for $\langle \bm{I}_{\rm Loc} \rangle$ and $\langle \bm{I}_{\rm Sur} \rangle$ are written as follows:
\begin{align}
    \frac{d \braket{ \bm I_{\rm Loc} }}{dt}  & = \frac{1}{T_{\rm eL}}[Q \braket{ \bm S_{\parallel} } - \braket{ \bm I_{\rm Loc} }]  \nonumber \\
                                             & - \frac{1}{T_{\rm R, Loc}} \braket{ \bm I_{\rm Loc} }
    - \frac{1-r}{T_{\rm LS}}[\braket{ \bm I_{\rm Loc} } - \braket{ \bm I_{\rm Sur} }],\label{eq5}                                             \\
    \frac{d \braket{ \bm I_{\rm Sur} } }{dt} & =  - \frac{r}{T_{\rm LS}}[\braket{ \bm I_{\rm Sur} } - \braket{\bm I_{\rm Loc} } ]
                                              - \frac{1}{T_{\rm R, Sur}} \braket{ \bm I_{\rm Sur} }, \label{eq6}                             \\
    \braket{ \bm I_{\parallel} }             & =  r\braket{ \bm I_{\rm Loc} } + (1-r)\braket{ \bm I_{\rm Sur} }.
    \label{eq7}
\end{align}
Here, $r$ ($0 \le r \le 1$) is the weight of the localization sites, and the sum of $r \langle \bm I_{\rm Loc} \rangle$ and $\left( 1-r \right) \langle \bm I_{\rm Sur} \rangle$ gives the observed NSP, $\langle \bm I_{\parallel} \rangle$. For simplicity, $1/T_{\rm eS}$ is assumed to be 0 in Eq.~\eqref{eq6} due to the weaker HFI for $\langle \bm{I}_{\rm Sur} \rangle$ (i.e., $T_{\rm eS} \gg T_{\rm eL}$). 

\begin{figure*}[t]
    \centering
    \includegraphics[width=\textwidth]{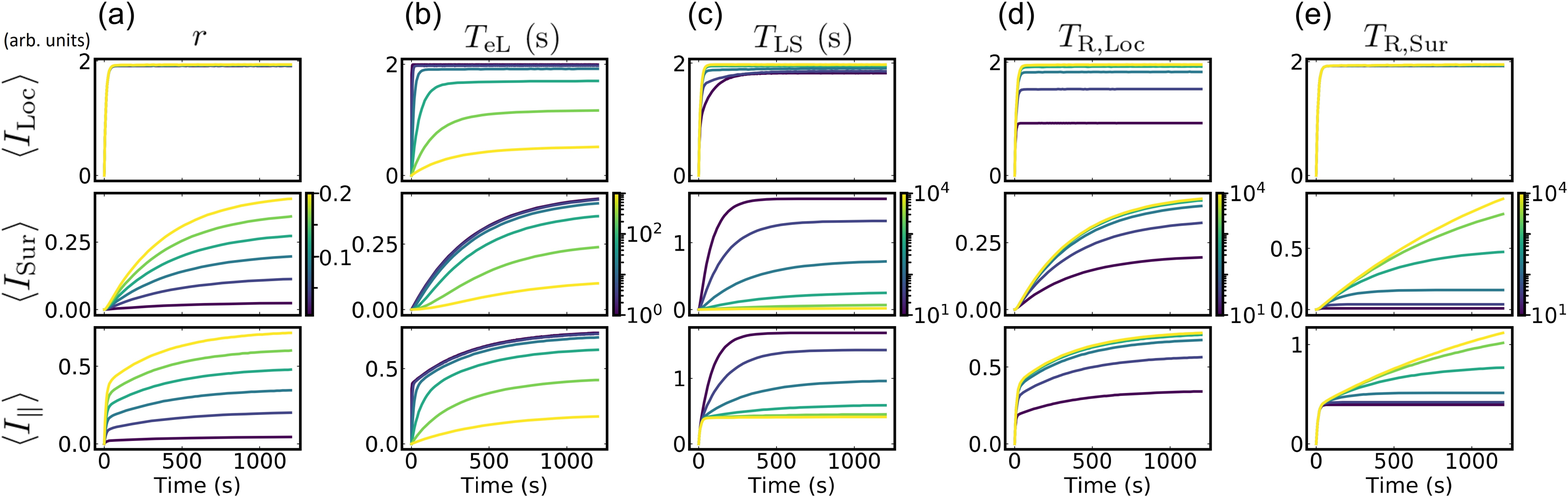}
    \caption{
        The calculations of formation dynamics of NSP under various parameter conditions.
        Each column shows the dependence on a single parameter shown at the top.
        From top to bottom, the rows display the dynamics of $\braket{I_{\rm Loc}}$, $\braket{I_{\rm Sur}}$, and $\braket{I_{\parallel}}$.
        When one parameter is swept, the others are fixed at their base values: $T_{\rm eL}=11.5$ s, $T_{\rm R, Loc}=1000$ s, $T_{\rm R, Sur}=550$ s, $T_{\rm LS}=300$ s, $r=0.2$. These values are appropriate for AlGaAs samples. The color bar indicates the value of each parameter.
    }
    \label{Fig5}
\end{figure*}

The spin transfer time from $\langle \bm{S}_{\parallel} \rangle$ to $\langle \bm{I}_{\rm Loc} \rangle$ is given as $T_{\rm eL} = 1/ (\bar{\Omega}_{\rm N}^2 \tau_{\rm c})$, where $\bar{\Omega}_{\rm N}$ is the average of nuclear Larmor frequency under the Knight field $B_{\rm e} \approx 10$ mT~\cite{PhysRevLett.111.087603, d1974optical, PhysRevB.15.5780}. $\tau_{\rm c}$ is the correlation time of localized electrons, which is estimated to be $\sim 10$ ps from the doping density of our samples~\cite{PhysRevB.66.245204}. Thus, $T_{\rm eL}$ is estimated to be $\sim 11.5$ s.

Note that the NSP relaxation rates, $1/T_{\rm R,Loc}$ and $1/T_{\rm R,Sur}$, consists of $B_{\rm ext}$-independent and $B_{\rm ext}$-dependent terms. The  former, denoted as $1/T_{\rm D}$, represents the nuclear spin diffusion. On the other hand, the latter one, $1/T_{\rm Q}$, originates from the fluctuations in the local magnetic fields $B_{\rm L}$ and is written as $1/T_{\rm Q} \propto 1/(B_{\rm ext}^2 + B_{\rm L}^2)$. This expression suggests that $1/T_{\rm Q}$ approaches zero under $\left| B_{\rm ext} \right| \gg B_{\rm L}$, resulting in longer $T_{\rm R, Loc (Sur)}$. As discussed below, the magnitude of $B_{\rm L}$ and the impact on $1/T_{\rm R, Loc (Sur)}$ depends on what the source of $B_{\rm L}$ is. Here, we only mention that different $1/T_{\rm R, Loc (Sur)}$ are assumed in the following calculations for strong and weak $B_{\rm ext}$ and for AlGaAs and GaAs.

\begin{table}[tb]
    \centering
    \caption{Used parameters for the calculations in Fig.~\ref{Fig4}(b).}
    \label{Table2}
    \begin{tabular}{c|c|ccccc}
    \hline
    materials & $B_{\rm ext}$ & $r$ & $T_{\rm LS}$ & $T_{\rm eL}$ & $T_{\rm R, Loc}$ & $T_{\rm R, Sur}$  \\
    \hline
    \multirow{2}{*}{AlGaAs} & $B_{\rm ext} \gg B_{\rm L}$ & \multirow{2}{*}{0.20} & \multirow{2}{*}{300 s} & \multirow{2}{*}{11.5 s} & 1000 s& 550 s\\ 
              & $B_{\rm ext}<B_{\rm L}$ &       &       &       & 10 s    & 30 s\\
    \hline          
    GaAs   & $B_{\rm ext} \gg B_{\rm L}$ & 0.01 & 300 s& 11.5 s & 1000 s& 130 s\\
    \hline
    \end{tabular}
\end{table}

Figure~\ref{Fig4}(b) summarizes the calculated results; the purple and brown curves correspond to AlGaAs and GaAs, respectively. The left (right) side represents the NSP formation (relaxation) dynamics, with $Q \langle S_{\parallel} \rangle$ set to $2.0$ ($0.0$). Further, the solid and dashed curves correspond to the one under strong and weak $B_{\rm ext}$, respectively. 
The other used parameters for the calculations are shown in Table~\ref{Table2}. 
In this model, NSP is generated only in the direction of $\bm{B}_{\rm ext}$. Hereafter, we basically focus  on the magnitudes of $\langle \bm{I}_{\rm Loc} \rangle$ and $\langle \bm{I}_{\rm Sur} \rangle$, that is, $\langle I_{\rm Loc} \rangle$ and $\langle I_{\rm Sur} \rangle$.

To understand the impact of each parameter on the NSP formation dynamics in this model, we examined the parameter dependences and summarized them in Fig.~\ref{Fig5}. For each parameter, the temporal evolutions of $\langle I_{\rm Loc} \rangle$ (top), $\langle I_{\rm Sur} \rangle$ (middle), and $\langle I_{\parallel} \rangle$ (bottom) were calculated.

\subsection{Overall features in NSP formation dynamics}
As shown on the left side of Fig.~\ref{Fig4}(b), our model successfully reproduced the experimentally observed two-stage formation dynamics, i.e., the rapid initial rise and the gradual increase in NSP. 

Figure~\ref{Fig5}(a) indicates that the key to explaining the two-stage dynamics is to divide the nuclear spin ensemble into two groups, i.e., the weight of the localization sites, $r$ must be other than $1$ and $0$. In this panel, $r$ was varied from 0.01 to 0.20. 
While $\langle I_{\rm Loc} \rangle$ reaches a unique saturation value after a rapid rise and is almost independent of $r$, the growth of $\langle I_{\rm Sur} \rangle$ is more pronounced when $r$ is large within this calculation range. Thus, as $r$ increases, the two-stage behavior becomes more prominent. 
It comes from the fact that the two groups, $\langle \bm{I}_{\rm Loc} \rangle$ and $\langle \bm{I}_{\rm Sur} \rangle$, develop with different characteristic times [$T^{-}$ and $T^{+}$ in Eqs.~\eqref{B10} and~\eqref{B11}] and the spin transfer rate between them is effectively proportional to $r$.
A single exponential behavior is obtained for $r=1$ and $0$, which is consistent with the expectation from the conventional model, Eq.~\eqref{eq4}.

Note that $\langle I_{\rm Loc} \rangle$ corresponds to the source of $B_{\rm f}^{\rm Form}$ and $\langle I_{\rm Sur} \rangle$ does $B_{\rm s}^{\rm Form}$ in our model. The parameters used in Fig.~\ref{Fig4}(b) were chosen so that $\left( 1-r \right) \langle I_{\rm Sur} \rangle / (r \langle I_{\rm Loc} \rangle )$ reproduces the experimentally evaluated $B_{\rm s}^{\rm Form}/ B_{\rm f}^{\rm Form}$.

The parameter $r$ is also important in explaining the differences in $B_{\rm s}^{\rm Form}/B_{\rm f}^{\rm Form}$ and $B_{\rm N}^{\rm max}$ between AlGaAs and GaAs. The experiments showed that these two quantities were larger for AlGaAs than for GaAs, which is consistent with the calculation if $r$ is larger for AlGaAs than for GaAs. 
The substitution of Ga by Al inevitably creates electron localization sites like charged impurities and defects, which may cause larger $r$ in AlGaAs.

Two-stage dynamics is explained also with the analytical solution of this model. 
As derived in Appendix~\ref{Append_B}, the solution of Eqs.~\eqref{eq5}-\eqref{eq7} confirms that the system exhibits a bi-exponential behavior with two characteristic time constants, $T_{\rm -}$ and $T_{\rm +}$, defined as follows: 
\begin{align} 
    \frac{1}{T_{-}} & \approx \frac{1}{T_{\rm eL}} + \frac{1}{T_{\rm R, Loc}} + \frac{1-r}{T_{\rm LS}}, \label{B10}\\
    \frac{1}{T_{+}} & \approx \frac{r}{T_{\rm LS}} + \frac{1}{T_{\rm R, Sur}}\label{B11}. 
\end{align}
$T_{-}$ represents the fast dynamics of $\langle \bm{I}_{\rm Loc} \rangle$ and $T_{+}$ does the slow one of $\langle \bm{I}_{\rm Sur} \rangle$, which are discussed in correspondence with $\tau_{\rm f}^{\rm Form}$ and $\tau_{\rm s}^{\rm Form}$, respectively.

$T_{-}$ and $T_{+}$ are determined by various time constants ($T_{\rm eL}$, $T_{\rm LS}$, $T_{\rm R, Lor}$, $T_{\rm R, Sur}$), but under strong magnetic fields ($\left| B_{\rm ext} \right| \gg B_{\rm L}$), $T_{\rm eL}$ which is the shortest, has a significant impact on NSP via the change in $T_{-}$. Figure~\ref{Fig5}(b) summarizes the impact of $T_{\rm eL}$. When $T_{\rm eL}$ becomes shorter (i.e., spin transfer from electrons occurs more efficiently), $\langle I_{\rm Loc} \rangle$ rises faster and reaches a larger value. Such $\langle I_{\rm Loc} \rangle$ leads to larger $\langle I_{\rm Sur} \rangle$, resulting in the appearance of larger $\langle I_{\parallel} \rangle$ with a clear two-stage structure.

Figure~\ref{Fig5}(c) summarizes the impact of $T_{\rm LS}$ which acts on both $T_{-}$ and $T_{+}$. The change in $T_{\rm LS}$ has only a slight change in rise time for $\langle I_{\rm Loc} \rangle$, but a marked impact on magnitude for $\langle I_{\rm Sur} \rangle$. As $T_{\rm LS}$ shortens, which means the efficient spin transfer from localization sites to surrounding area, $\langle I_{\rm Loc} \rangle$ rises slower while $\langle I_{\rm Sur} \rangle$ rises faster to a larger saturation value. Further, when $T_{\rm LS}$ is as short as or shorter than $T_{\rm eL}$, the difference between $T_{-}$ and $T_{+}$ becomes smaller and the dynamics of $\langle I_{\parallel} \rangle$ approaches a single exponential behavior. This suggests that frequent spin exchanges between $\langle {I}_{\rm Loc} \rangle$ and $\langle {I}_{\rm Sur} \rangle$ make the entire system homogeneous on a short timescale.

Since $T_{\rm R, Loc}$ and $T_{\rm R, Sur}$ vary widely depending on which mechanism dominates the nuclear spin relaxation, their impacts on NSP are discussed in the next subsection. 
From Eqs.~\eqref{B10} and~\eqref{B11}, we estimated $T_{-} \sim 11$ s (common for AlGaAs and GaAs), $T_{+} \sim 400$ s (for AlGaAs), and $T_{+} \sim 130$ s (for GaAs), respectively. Although $\tau_{\rm f}^{\rm Form}$ to be compared to $T_{-}$ has not been measured yet, the estimated value of $T_{+}$ is close to $\tau_{\rm s}^{\rm Form}$ for each sample (the average of samples A and B for AlGaAs).

\subsection{Impact of NSP relaxation rates and NQI}

Here, we investigate how nuclear spin relaxation rates, $1/T_{\rm R, Loc}$ and $1/T_{\rm R, Sur}$, vary with $B_{\rm ext}$ and sample, and how they affect the NSP dynamics.

As mentioned earlier, $1/T_{\rm R, Loc}$ and $1/T_{\rm R, Sur}$ consist of $B_{\rm ext}$-dependent ($1/T_{\rm Q}$) and $B_{\rm ext}$-independent ($1/T_{\rm D}$) terms, and the magnitude of $1/T_{\rm Q}$ and the range over which it works are determined by the local field, $B_{\rm L}$. The possible sources of $B_{\rm L}$ are the non-secular terms of the dipole interaction among nuclei (${\cal H}_{\rm dd}$) and NQI~\cite{note1}. For strain-free GaAs samples, the reported value of $B_{\rm L}$ is a few Gausses~\cite{kotur2016nuclear, vladimirova2017nuclear}. Although the strength of ${\cal H}_{\rm dd}$ is comparable, the reported values of $B_{\rm L}$ for ternary compound semiconductors range from a few mT to several tens of mT, even for strain-free samples~\cite{rittmann2022unveiling}. This is due to the additional contribution of NQI with larger strength, that is, NQI-induced relaxation~\cite{kotur2016nuclear}.

In Fig.~\ref{Fig3}(c), $B_{\rm N}$ in samples A and B was nearly constant as $B_{\rm ext} \gtrsim 50$ mT, while it decreased with decreasing $B_{\rm ext}$ in the weaker range. It indicated that $B_{\rm L}$ in AlGaAs bulk is about a few ten mT. In contrast, the $B_{\rm N}$ of GaAs was nearly constant at $\left| B_{\rm ext} \right| \geq 6$ mT, indicating that the $B_{\rm L}$ of GaAs is well below 6 mT. These results are consistent with previous reports on $B_{\rm L}$. We also found that the effects of $1/T_{\rm Q}$ were well suppressed at 560 mT, where we conducted experiments.

We now describe the calculation results for the NSP relaxation dynamics shown in Fig.~\ref{Fig4}(b). For AlGaAs, $T_{\rm R, Loc}$ and $T_{\rm R, Sur}$ were assumed to be 1000 s and 550 s for strong-$B_{\rm ext}$ situation, and shorter values ($T_{\rm R, Loc} =$ 10 s and $T_{\rm R, Sur} =$ 30 s) for weak-$B_{\rm ext}$. Using these values, noticeable differences appear between the strong (purple solid curve) and weak (dashed curve) $B_{\rm ext}$. For strong-$B_{\rm ext}$, $\langle I_{\parallel} \rangle$ decreases to about half of its initial value on short time ($\sim 10$ s) and then approaches zero on long timescale ($\sim 100$ s). On the other hand, $\langle I_{\parallel} \rangle$ for weak-$B_{\rm ext}$ relaxes to zero in a short timescale of $\sim 10$ s.

Fast relaxation for strong-$B_{\rm ext}$, which comes from the term $1/T_{\rm eL}$ with $\langle S_{\parallel} \rangle =0$, was not detected due to the limitation of the measurement. However, the calculated results are consistent with the experimental results in that more intense relaxation occurs under weak $B_{\rm ext}$ than under strong $B_{\rm ext}$. 
In the experiment, only a slow relaxation component ($\tau_{\rm s}^{\rm Relax} =248$ s and 306 s) was observed at $+560$ mT, while an additional fast component ($\tau_{\rm f}^{\rm Relax} \sim 30$ s) with significant amplitude ($B_{\rm f}^{\rm Relax} = 279$ mT and 303 mT) appeared at 6 mT.

Another impact of NQI is the suppression of nuclear spin diffusion. NQI makes nuclear spin state splitting unequal and inhibits the spin flip-flops between neighboring nuclei~\cite{millington2023nuclear, maletinsky2007dynamics, adachi2012nuclear}. In single QDs, this suppression effect remains even under $B_{\rm ext}$ of a few tesla~\cite{maletinsky2009breakdown}. Based on this, the calculation assuming $T_{\rm R, Sur} = 130$ s for GaAs in a strong $B_{\rm ext}$, which was shorter than for AlGaAs (but $T_{\rm R, Loc}$ was the same as for AlGaAs), yielded a shorter relaxation time of slow component than for AlGaAs. This is consistent with the experimental result that $\tau_{\rm s}^{\rm Relax}$ at a strong $B_{\rm ext}$ was shorter for GaAs than for AlGaAs (more pronounced for sample B).

Here, we consider the validities of adopting a longer $T_{\rm R, Loc}$ than $T_{\rm R, Sur}$ and using the same $T_{\rm R, Loc}$ for AlGaAs and GaAs in calculations for strong $B_{\rm ext}$. Note that in this model, we assume that nuclear spin diffusion is the main mechanism for the spin transfer between $\langle \bm{I}_{\rm Loc} \rangle$ and $\langle \bm{I}_{\rm Sur} \rangle$. 
Nuclear spin diffusion that remains even under strong $B_{\rm ext}$ contributes mainly to spin transfer to $\langle \bm{I}_{\rm Sur} \rangle$ in the localization sites, but in the surrounding region, 
it does not only to spin transfer to $\langle \bm{I}_{\rm Loc} \rangle$ but also to leakage outside the observation area. In other words, $1/T_{\rm R, Sur}$ includes the effect of spin diffusion, but $1/T_{\rm R, Loc}$ does not (instead, that effect is included in $1/T_{\rm LS}$). So, it is reasonable to assume that $T_{\rm R, Loc}$ is longer than $T_{\rm R, Sur}$ in strong $B_{\rm ext}$. Furthermore, it is also reasonable that there is no difference between AlGaAs and GaAs in $T_{\rm R, Loc}$ when the relaxation mechanism modulated by NQI is suppressed, as with $\langle \bm{I}_{\rm Loc} \rangle$ under strong $B_{\rm ext}$.

Figure~\ref{Fig5}(d) and \ref{Fig5}(e) show the effect of $1/T_{\rm R, Loc}$ and $1/T_{\rm R, Sur}$ on NSP formation dynamics, respectively. In both panels, a larger $\langle I_{\parallel} \rangle$ is created with a longer nuclear spin relaxation time (i.e., a smaller relaxation rate). This is consistent with the intuitive expectation that smaller leakage from the system is more favorable for larger $B_{\rm N}$ formation, and with experimental results showing larger $B_{\rm N}$ under strong $B_{\rm ext}$, where $1/T_{\rm Q}$ is suppressed and relaxation times are extended. However, the effects of $T_{\rm R, Loc}$ and $T_{\rm R, Sur}$, especially on $\langle I_{\rm Loc} \rangle$, are different.

$T_{\rm R, Loc}$ (in the range of 10 to 10$^{4}$ s) affects the saturation value of $\langle I_{\rm Loc} \rangle$ as well as $T_{\rm eL}$, but there is no significant change in $T_{-}$, the rise time of $\langle I_{\rm Loc} \rangle$. This indicates that $T_{-}$ is determined almost exclusively by $T_{\rm eL}$, whereas the saturation value of $\langle I_{\rm Loc} \rangle$ is determined by the balance between pumping from ESP ($1/T_{\rm eL}$) and relaxation ($1/T_{\rm R, Loc}$). On the other hand, $\langle I_{\rm Loc} \rangle$ hardly changes when $T_{\rm R, Sur}$ changes. This is because we assumed in Fig.~\ref{Fig5}(e) that $T_{\rm eL}$ was quite shorter than $T_{\rm R, Loc}$ and $T_{\rm LS}$, i.e., the small leakage of $\langle I_{\rm Loc} \rangle$. Therefore, even if the leakage of $\langle \bm{I}_{\rm Sur} \rangle$ is large, little impact on $\langle \bm{I}_{\rm Loc} \rangle$ occurs.

Finally the nuclear spin diffusion coefficients $D$ in GaAs and AlGaAs bulks are estimated. If $T_{\rm R, Sur}$ under a strong $B_{\rm ext}$ is regarded as $T_{\rm D}$ due to the suppression of $1/T_{\rm Q}$, $D$ is denoted as $\displaystyle D = {(4 \pi a n_{\rm d}T_{\rm D})}^{-1}$~\cite{de1958relaxation}. Here, $a$ is the localization radius of electrons and is about 10 nm~\cite{kotur2016nuclear}. 
From calculations, $D$ was estimated to be $\approx 5.2 \times 10^{-14}$ ${\rm cm}^2/{\rm s}$ for sample C and $\approx 1.3 \times 10^{-14}$ ${\rm cm}^2/{\rm s}$ in samples A and B. The value for sample C was smaller than the reported one for GaAs bulk~\cite{PhysRevB.25.4444}. Furthermore, it was found that the estimated value for AlGaAs was smaller than the one for GaAs, suggesting that nuclear spin diffusion is well suppressed by NQI.



\section{Conclusion}

In this study, 
we investigated the formation and relaxation dynamics of NSP in bulk $n$-doped Al$_x$Ga$_{1-x}$As with different aluminum concentration $x=$ 0.15, 0.05, and 0.00 by using the TRKR technique.

After determining the magnitude, sign, and anisotropy of the electron g-factor ${\rm g}_{\rm e}$ for the three samples, the  ${\bm B}_{\rm N}$ was used as a measure of NSP, and the erase-prepump-(dark-)pump/probe method was used to measure the formation and relaxation processes of the NSP. 
The results showed that ${\bm B}_{\rm N}$ had an anomalously large magnitude and a long formation time, which involved an apparent two components with different formation rates. 

Relaxation measurements showed that AlGaAs samples had a slower relaxation of ${\bm B}_{\rm N}$ than the GaAs sample. On the other hand, the comparison between results taken under strong and weak $B_{\rm ext}$ indicated the suppression of NQI-induced relaxation.

We updated the NSP formation model and simulated the formation and relaxation dynamics based on the model. Calculations under different parameters were also carried out. The calculated results were consistent with our experimental results.

\begin{acknowledgments}
    We acknowledge the support of the Japan Society for the Promotion of Science (JSPS) KAKENHI Grant Numbers 
JP24K08189 and JP25K01604. This work was partially supported by a research grant from The Mazda Foundation and JST SPRING, Grant Number JPMJSP2119.
\end{acknowledgments}

\appendix
\section{Nuclear field generation in oblique magnetic field and g-factor isotropy} \label{Append_A}

\begin{figure}[htb]
    \centering
    \includegraphics[width=240 pt]{Fig1A.eps}
    \caption{
    (a) Diagram of $\bm{B}_{\rm N}$ generation under oblique ${\bm B}_{\rm ext}$ with positive ${\rm g}_{\rm e}$. 
    Blue arrows indicate the ESP injected by $\sigma^{\pm}$ pump light. The gray and red arrows are ${\bm B}_{\rm ext}$ and ${\bm B}_{\rm N}$.  (b)  $\nu_{\rm L}$-dependence of $B_{\rm ext}$ of sample A measured with different $\theta_{\rm B}$. The $\nu_{\rm L}$ of each data set is shifted for easy viewing. The inset shows the corresponding $|{\rm g}_{\rm e}|$ dependence of $\theta_{\rm B}$. (c) Dependence of TRKR signals of sample A on $\theta_{\rm B}$ ranging from -18$^\circ$ to +18$^\circ$ at $B_{\rm ext} = +560$ mT without (left) and with (right) ${\bm B}_{\rm N}$. The bottom panels are the corresponding Fourier spectra. (d) Measured value of $\nu_{\rm L}$ as a function of $\theta_{\rm B}$. The right axis indicates the corresponding $B_{\rm N}$ value. 
    }
    \label{Fig1A}
\end{figure}

Under the oblique magnetic field, the generation of the nuclear field $\bm{B}_{\rm N}$ occurs~\cite{OptOrientation84, SpinPhysicsinSC}.  
As shown in Fig.~\ref{Fig1A}(a), the electron spin precessing around an oblique magnetic field is divided into two components, $\braket{S_{\parallel}}$ parallel to $\bm{B}_{\rm ext}$ and $\braket{S_{\perp}}$ perpendicular to $\bm{B}_{\rm ext}$. $\braket{S_{\perp}}$ precess around $\bm{B}_{\rm ext}$ while $\braket{S_{\parallel}}$ remained its direction along the $\bm{B}_{\rm ext}$. As a result, the nuclear spin is polarized by $\braket{S_{\parallel}}$ through collinear HFI and generates $\bm{B}_{\rm N}$.
Depending on the pump helicity, $\bm{B}_{\rm N}$ is generated parallel or antiparallel to $\bm{B}_{\rm ext}$, which is reflected in the increase or decrease of the $\nu_{\rm L}$, as shown in Figs.~\ref{Fig2}(c) and~\ref{Fig2}(d).

To check the ${\rm g}_{\rm e}$ isotropy, the experiments similar to Fig.~\ref{Fig2}(b) were conducted in other $\theta_{\rm B}$ by using a PEM. The $\nu_{\rm L}$-dependence on $B_{\rm ext}$ with different $\theta_{\rm B}$ is shown in Fig.~\ref{Fig1A}(b).
For ease of viewing, $\nu_{\rm L}$ is plotted in the range from -0.6 GHz to 0.6 GHz, shifted at 0.2 GHz intervals around $\theta_{\rm B} = 0^{\circ}$.
All data show the same trend, and the ${\rm g}_{\rm e}$ obtained at different $\theta_{\rm B}$ by fitting using Eq.~\eqref{eq1}  is shown in the inset of Fig.~\ref{Fig1A}(b). The minimal difference indicated no significant anisotropy of the ${\rm g}_{\rm e}$, which coincided with the fact that the g-factor in zinc blende bulk semiconductors was isotropic~\cite{Shchepetilnikov_2013}.

Next, the EOM was applied to generate the $\bm{B}_{\rm N}$ under an oblique $\bm{B}_{\rm ext}$. Figure~\ref{Fig1A}(c) shows 2D plots of the $\theta_{\rm B}$-dependence of TRKR signals of sample A without ${\bm B}_{\rm N}$ (left, by PEM) and with ${\bm B}_{\rm N}$ (right, $\sigma^+$ excitation by EOM), respectively.
$B_{\rm ext}$ was fixed at $+560$ mT and $\theta_{\rm B}$ was varied in the range from $-18^{\circ}$ to $+18^{\circ}$.
The Fourier-transformed spectra of the TRKR signals are also shown in the bottom panels.

In the left of Fig.~\ref{Fig1A}(c), $\nu_{\rm L}$ changed minimally with respect to $\theta_{\rm B}$, but the initial phase $\phi$ changed significantly when PEM was used.
The initial phase was determined by the formation dynamics of RESP~\cite{PhysRevB.90.205307}. Our observation suggested that the orientation of the RESP changed with $\theta_{\rm B}$ upon the return of the captured electron from the excitons to the electron ensemble.
However, in the upper right panel, the dependency of $\nu_{\rm L}$ on $\theta_{\rm B}$ was observed with EOM.
This variation was clear in the Fourier-transformed spectra (bottom right panel), where the peak position was apparently changed.
The slight change in ${\rm g}_{\rm e}$ with $\theta_{\rm B}$ shown in the inset of Fig.~\ref{Fig1A}(b) cannot account for this difference in $\nu_{\rm L}$, and therefore, a relatively large $B_{\rm N}$ must have been generated.
In addition, compared to the EOM case, PEM suppressed the change of $\nu_{\rm L}$ with $\theta_{\rm B}$, which provides another proof of the generation of $B_{\rm N}$.

Figure~\ref{Fig1A}(d) shows a clearer comparison of $\nu_{\rm L}$ obtained with and without ${\bm B}_{\rm N}$ in Fig.~\ref{Fig1A}(c).
The corresponding $B_{\rm N}$, calculated by $h \nu_{\rm L}/(\mu_{\rm B}|{\rm g}_{\rm e}| ) - B_{\rm ext}$, is shown on the right axis.
The black circles represent the data taken with PEM, indicating almost no $B_{\rm N}$ was generated, while the red squares represent the data taken with the EOM, indicating that the $|B_{\rm N}|$ became larger with increasing $|\theta_{\rm B}|$, as $\braket{S_{\parallel}}$ in Eq.~\eqref{eq4} equals $\braket{S_0}{\rm sin}\theta_{\rm B}$.
A certain amount of $B_{\rm N}$ at $\theta_{\rm B} = 0^{\circ}$ is the $B_{\rm N}$ remaining from the preceding measurement, suggesting a slow relaxation of the NSP.

\section{Analytical solution of the rate equations in model calculations} \label{Append_B}

Here, we present the analytical solution of the rate equations used for model calculations of the NSP formation and relaxation dynamics.
First, to solve the rate equations for $\braket{\bm I_{\rm Loc}}$, and $\braket{\bm I_{\rm Sur}}$, as presented in Eqs.~\eqref{eq5} and \eqref{eq6}, we rewrite the equations as
\begin{align}
    \frac{d\braket{\bm I_{\rm Loc}}}{dt} & = -\frac{1}{T_{\rm Loc}}\braket{\bm I_{\rm Loc}} + \frac{1-r}{T_{\rm LS}}\braket{\bm I_{\rm Sur}} + \frac{Q\braket{\bm S_{\parallel}}}{T_{\rm eL}}, \label{eqb1} \\
    \frac{d\braket{\bm I_{\rm Sur}}}{dt} & = \frac{r}{T_{\rm LS}}\braket{\bm I_{\rm Loc}} - \frac{1}{T_{\rm Sur}}\braket{\bm I_{\rm Sur}},\label{eqb2}
\end{align}
where the effective time constants $T_{\rm Loc}$ and $T_{\rm Sur}$ are

\begin{align}
    \frac{1}{T_{\rm Loc}} & = \frac{1}{T_{\rm eL}} + \frac{1}{T_{\rm R, Loc}} + \frac{1-r}{T_{\rm LS}}, \label{eqb3} \\
    \frac{1}{T_{\rm Sur}} & = \frac{r}{T_{\rm LS}} + \frac{1}{T_{\rm R, Sur}}. \label{eqb4}
\end{align}

The general solution to this system is the sum of a particular (steady-state) solution and a homogeneous solution. The steady state of the system ($\frac{dI}{dt} = 0$) is given by
\begin{align}
    \braket{\bm I_{\rm Loc}^{\rm stdy}} & = \frac{T_{\rm Loc} T_{\rm LS}^2}{T_{\rm LS}^2 - r(1-r)T_{\rm Loc}T_{\rm Sur}} \frac{Q\braket{\bm S_{\parallel}}}{T_{\rm eL}}, \label{eqb5}            \\
    \braket{\bm I_{\rm Sur}^{\rm stdy}} & = \frac{r T_{\rm Loc} T_{\rm LS} T_{\rm Sur}}{T_{\rm LS}^2 - r(1-r)T_{\rm Loc}T_{\rm Sur}} \frac{Q\braket{\bm S_{\parallel}}}{T_{\rm eL}}.\label{eqb6}
\end{align}
The time-dependent part of the solution is governed by the eigenvalues of the system, which determine the characteristic relaxation rates:
\begin{align}
    -\frac{1}{T_{\pm}} = & -\frac{1}{2T_{\rm Loc}} - \frac{1}{2T_{\rm Sur}} \nonumber                                                           \\
                         & \pm \sqrt{\left(\frac{1}{2T_{\rm Loc}} - \frac{1}{2T_{\rm Sur}}\right)^2 + \frac{r(1-r)}{T_{\rm LS}^2}}.\label{eqb7}
\end{align}
\noindent Therefore, the full general solution, describing the evolution from an initial state $\braket{\bm I(t = 0)}$ to the steady state $\braket{\bm I^{\rm stdy}}$, is written as a double exponential function of time $t$:
\begin{align}
    \braket{\bm I_{\rm Loc}(t)} & = C_{11}e^{-t/T_{-}} + C_{12}e^{-t/T_{+}} + \braket{\bm I_{\rm Loc}^{\rm stdy}}, \label{eqb8} \\
    \braket{\bm I_{\rm Sur}(t)} & = C_{21}e^{-t/T_{-}} + C_{22}e^{-t/T_{+}} + \braket{\bm I_{\rm Sur}^{\rm stdy}}. \label{eqb9}
\end{align}
\noindent The coefficients $C_{ij}$ depend on the initial conditions and system parameters. Both $\braket{\bm I_{\rm Loc}(t)}$ and $\braket{\bm I_{\rm Sur}(t)}$ share the same two characteristic time constants, $T_+$ and $T_-$, which govern the fast and slow components of the system's dynamics.

Assuming that $T_{\rm eL}$ is much smaller than $T_{\rm R, Loc}$, $T_{\rm R, Sur}$, and $T_{\rm LS}$, the following approximation for the eigenvalues is obtained:
\begin{align}
    \frac{1}{T_{-}} & \approx \frac{1}{T_{\rm eL}} + \frac{1}{T_{\rm R, Loc}} + \frac{1-r}{T_{\rm LS}} = \frac{1}{T_{\rm Loc}}, \label{eqb10}\\
    \frac{1}{T_{+}} & \approx \frac{r}{T_{\rm LS}} + \frac{1}{T_{\rm R, Sur}}  = \frac{1}{T_{\rm Sur}}.\label{eqb11}
\end{align}
\noindent 
This Eq.~\eqref{eqb10} suggests that $T_-$, which describes a fast evolution, is mainly determined by $T_{\rm eL}$. In contrast, other exponential relaxation characterized by $T_+$ describes a slow evolution determined by a competition of $T_{\rm R, Sur}$, and $T_{\rm LS}$. The dependence on these time constants is well shown in Fig.~\ref{Fig5}.


\begin{thebibliography}{99}
    \bibitem{awschalom2002semiconductor} D. Awschalom, D. Loss, and N. Samarth, \textit{Semiconductor spintronics and quantum computation} (Springer Science \& Business Media, 2002).

    \bibitem{xu2023two} H. Xu, C. Li, G. Wang, H. Wang, H. Tang, A. R. Barr, P. Cappellaro, and J. Li, Two-photon interface of nuclear spins based on the optonuclear quadrupolar effect, Phys. Rev. X \textbf{13}, 011017 (2023).

    \bibitem{uysal2023coherent} M. T. Uysal, M. Raha, S. Chen, C. M. Phenicie, S. Ourari, M. Wang, C. G. Van de Walle, V. V. Dobrovitski, and J. D. Thompson, Coherent control of a nuclear spin via interactions with a rare-earth ion in the solid state, PRX Quantum \textbf{4}, 010323 (2023).

    \bibitem{pla2013high} J. J. Pla, K. Y. Tan, J. P. Dehollain, W. H. Lim, J. J. Morton, F. A. Zwanenburg, D. N. Jamieson, A. S. Dzurak, and A. Morello, High-fidelity readout and control of a nuclear spin qubit in silicon, Nature \textbf{496}, 334 (2013).

    \bibitem{gangloff2019quantum} D. Gangloff, G. \'{E}thier-Majcher, C. Lang, E. Denning, J. Bodey, D. Jackson, E. Clarke, M. Hugues, C. Le Gall, and M. Atat\"{u}re, Quantum interface of an electron and a  nuclear ensemble, Science \textbf{364}, 62 (2019).

    \bibitem{bodey2019optical} J. Bodey, R. Stockill, E. Denning, D. Gangloff, G. \'{E}thier-Majcher, D. Jackson, E. Clarke, M. Hugues, C. L. Gall, and M. Atat\"{u}re, Optical spin locking of a solid-state qubit, npj Quantum Inf. \textbf{5}, 95 (2019).

    \bibitem{gangloff2021witnessing} D. A. Gangloff, L. Zaporski, J. H. Bodey, C. Bachorz, D. M. Jackson, G. \'{E}thier-Majcher, C. Lang, E. Clarke, M. Hugues, C. Le Gall, and M. Atat\"{u}re, Witnessing quantum correlations in a nuclear ensemble via an electron spin qubit, Nat. Phys. \textbf{17}, 1247 (2021).

    \bibitem{jackson2021quantum} D. M. Jackson, D. A. Gangloff, J. H. Bodey, L. Zaporski, C. Bachorz, E. Clarke, M. Hugues, C. Le Gall, and M. Atat\"{u}re, Quantum sensing of a coherent single spin excitation in a nuclear ensemble, Nat. Phys. \textbf{17}, 585 (2021).

    \bibitem{mocek2017high}R. W. Mocek, V. L. Korenev, M. Bayer, M. Kotur, R. I. Dzhioev, D. O. Tolmachev, G. Cascio, K. V. Kavokin, and D. Suter, High-efficiency optical pumping of nuclear polarization in a GaAs quantum well, Phys. Rev. B \textbf{96}, 201303 (2017).

    \bibitem{urbaszek2013nuclear} B. Urbaszek, X. Marie, T. Amand, O. Krebs, P. Voisin, P. Malentinsky, A. H\"{o}gele, and A. Imamoglu, Nuclear spin physics in quantum dots: An optical investigation, Rev. Mod. Phys. \textbf{85}, 79 (2013).

    \bibitem{PhysRevB.101.245424} S. Yamamoto, R. Kaji, H. Sasakura, and S. Adachi, Third stable branch and tristability of nuclear spin polarizations in a single quantum dot system, Phys. Rev. B \textbf{101}, 245424 (2020).

    \bibitem{huang2020oblique} Y. Q. Huang, Y. Puttisong, S. Filippov, I. A. Buyanova, and W. M. Chen, Oblique nuclear quadrupole interaction in self-assembled structures based on semiconductor quantum dots, Phys. Rev. Applied \textbf{14}, 044019 (2020).

    \bibitem{Kreb_PRL2010} O. Krebs, P. Maletinsky, T. Amand, B. Urbaszek, A. Lema\^{i}tre, P. Voisin, X. Marie, and A. Imamoglu, Anomalous Hanle Effect due to Optically Created Transverse Overhauser Field in Single InAs/GaAs Quantum Dots, Phys. Rev. Lett. \textbf{104}, 056603 (2010).

    \bibitem{Yamamoto_PRB18} S. Yamamoto, R. Matsusaki, R. Kaji, and S. Adachi, In-plane nuclear field formation investigated in single self-assembled quantum dots, Phys. Rev. B \textbf{97}, 075309 (2018).

    \bibitem{Yamamoto_PRB23} S. Yamamoto, R. Kaji, and S. Adachi, Anomalous hanle effect considered in time-resolved measurements and numerical simulations, Phys. Rev. B \textbf{108}, 054422 (2023).

    \bibitem{Latta_NatPhys2009} C. Latta, A. Hogele, Y. Zhao, A. N. Vamivakas, P. Maletinsky, M. Kroner, J. Dreiser, I. Carusotto, A. Badolato, D. Schuh,W. Wegscheider, M. Atature, and A. Imamoglu, Confluence of resonant laser excitation and bidirectional quantum-dot nuclear-spin polarization, Nat. Phys. \textbf{5}, 758-763 (2009).

    \bibitem{Hogele_PRL2012} A. H\"{o}gele, M. Kroner, C. Latta, M. Claassen, I. Carusotto, C. Bulutay, and A. Imamoglu, Dynamic Nuclear Spin Polarization in the Resonant Laser Excitation of an InGaAs Quantum Dot, Phys. Rev. Lett. \textbf{108}, 197403 (2012).

    \bibitem{ihm1980valence} J. Ihm, M. L. Cohen, and V. Safarov, Valence charge distribution and electric field gradients in GaAs-AlAs mixed crystals, Solid State Commun. \textbf{34}, 325 (1980).

    \bibitem{zakharchenya1981optical} B. Zakharchenya, V. Kalevich, V. Kul'kov, and V. Fleisher, Optical orientation of electron-nuclear spin-system of semiconductor in an inclined magnetic field, Fiz. Tverd. Tela \textbf{23}, 1387 (1981).

    \bibitem{OptOrientation84} \textit{Optical Orientation}, Modern Problems in Condensed Matter Sciences Vol. 8, edited by F. Meier and B. Zakharchenya (North-Holland, NewYork, 1984).



    \bibitem{artemova1991sinergetic} E. Artemova, E. Galaktionov, V. Kalevich, V. Korenev, I. Merkulov, and A. Silbergleit, Sinergetic phenomena in the electron-nucleus spin system of AlGaAs in a strong magnetic field, Nonlinearity \textbf{4}, 49 (1991).



    \bibitem{SNS1} V. S. Zapasskii, Spin-noise spectroscopy: from proof of principle to applications, Adv. Opt. Photon. \textbf{5}, 131 (2013).

    \bibitem{SNS2} M. Oestreich, M. R\"{o}mer, R. J. Haug, and D. H\"{a}gele, Spin noise spectroscopy in GaAs, Phys. Rev. Lett.  \textbf{95}, 216603 (2005).

    \bibitem{SNS3} S. A. Crooker, L. Cheng, and D. L. Smith, Spin noise of conduction electrons in n-type bulk GaAs, Phys. Rev. B \textbf{79}, 035208 (2009).



    \bibitem{vladimirova2017nuclear}M. Vladimirova, S. Cronenberger, D. Scalbert, M. Kotur, R. I. Dzhioev, I. I. Ryzhov, G. G. Kozlov, V. S. Zapasskii, A. Lema\^{i}tre, and K. V. Kavokin, Nuclear spin relaxation in n-GaAs: From insulating to metallic regime, Phys. Rev. B. \textbf{95}, 125312 (2017).

    \bibitem{vladimirova2018spintemp}M. Vladimirova, S. Cronenberger, D. Scalbert, I. I. Ryzhov, V. S. Zapasskii, G. G. Kozlov, A. Lema\^{i}tre, and K. V. Kavokin, Spin temperature concept verified by optical magnetometry of nuclear spins, Phys. Rev. B. \textbf{97}, 041301(R) (2018).

    \bibitem{QDlaser} QD Laser Inc., \url{https://www.qdl-laser.com/en/quantum-dot-laser-wafer}.




    \bibitem{vurgaftman2001band} I. Vurgaftman, J. Meyer, and L. R. Ram-Mohan, Band parameters for III--V compound semiconductors and their alloys, J. Appl. Phys. \textbf{89}, 5815 (2001).

    \bibitem{PhysRevB.66.245204}  R. I. Dzhioev, K. V. Kavokin, V. L. Korenev, M. V. Lazarev, B. Y. Meltser, M. N. Stepanova, B. P. Zakharchenya, D. Gammon, and D. S. Katzer, Low-temperature spin relaxation in $n$-type GaAs, Phys. Rev. B \textbf{66}, 245204 (2002).

    \bibitem{baumberg1994ultrafast} J. J. Baumberg, S. A. Crooker, D. D. Awschalom, N. Samarth, H. Luo, and J. K. Furdyna, Ultrafast faraday spectroscopy in magnetic semiconductor quantum structures, Phys. Rev. B \textbf{50}, 7689 (1994).

    \bibitem{crooker1997optical} S. A. Crooker, D. D. Awschalom, J. J. Baumberg, F. Flack, and N. Samarth, Optical spin resonance and transverse spin relaxation in magnetic semiconductor quantum wells, Phys. Rev. B \textbf{56}, 7574 (1997).

    \bibitem{Kikkawa1997} J. M. Kikkawa, I. P. Smorchkova, N. Samarth, and D. D. Awschalom, Room-temperature spin memory in two-dimensional electron gases, Science \textbf{277}, 1284 (1997).




    \bibitem{kennedy2006optical} T. A. Kennedy, A. Shabaev, M. Scheibner, A. L. Efros, A. S. Bracker, and D. Gammon, Optical initialization and dynamics of spin in a remotely doped quantum well, Phys. Rev. B \textbf{73}, 045307 (2006).


    \bibitem{colton2002spin} J. Colton, T. Kennedy, A. Bracker, and D. Gammon, Spin lifetime measurements in MBE-grown GaAs epilayers, phys. stat. sol. (b) \textbf{233}, 445 (2002).

    \bibitem{SpinPhysicsinSC} \textit{Spin Physics in Semiconductors}, Springer Series in Solid-State Sciences 157, edited by M. I. Dyakonov (Springer-Verlag, Berlin Heidelberg, 2008).

    \bibitem{PhysRevLett.80.4313} J. M. Kikkawa and D. D. Awschalom, Resonant spin amplification in $\mathit{n}$-type GaAs, Phys. Rev. Lett. \textbf{80}, 4313 (1998).

    \bibitem{PhysRevB.85.121303} A. Greilich, A. Pawlis, F. Liu, O. A. Yugov, D. R. Yakovlev, K. Lischka, Y. Yamamoto, and M. Bayer, Spin dephasing of fluorine-bound electrons in ZnSe, Phys. Rev. B \textbf{85}, 121303 (2012).



    \bibitem{chekhovich2013element} E. Chekhovich, M. Glazov, A. Krysa, M. Hopkinson, P. Senellart, A. Lema\^{i}tre, M. Skolnick, and A. Tartakovskii, Element-sensitive measurement of the hole--nuclear spin interaction in quantum dots, Nat. Phys. \textbf{9}, 74 (2013).

    \bibitem{weisbuch1977optical} C. Weisbuch and C. Hermann, Optical detection of conduction-electron spin resonance in GaAs, Ga$_{1- x}$In$_{x}$As, and Ga$_{1- x}$Al$_{x}$As, Phys. Rev. B \textbf{15}, 816 (1977).

    \bibitem{chadi1976gamma} D. J. Chadi, A. H. Clark, and R. D. Burnham, $\Gamma_1$ conduction electron g factor and matrix elements in GaAs and Al$_{x}$Ga$_{1- x}$As alloys, Phys. Rev. B \textbf{13}, 4466 (1976).

    \bibitem{yugova2007universal} I. A. Yugova, A. Greilich, D. R. Yakovlev, A. A. Kiselev, M. Bayer, V. V. Petrov, Y. K. Dolgikh, D. Reuter, and A. D. Wieck, Universal behavior of the electron g factor in GaAs/Al$_{x}$Ga$_{1-x}$As quantum wells, Phys. Rev. B \textbf{75}, 245302 (2007).

    \bibitem{malinowski2000nuclear} A. Malinowski and R. Harley, Dynamic nuclear Overhauser shifts in Larmor beats from a quantum well, Solid State Commun. \textbf{114}, 419 (2000).

    \bibitem{malinowski2001nuclear} A. Malinowski, M. Brand, and R. Harley, Nuclear effects in ultrafast quantum-well spin-dynamics, Physica E \textbf{10}, 13 (2001).

 \bibitem{PhysRevLett.111.087603} R. Giri, S. Cronenberger, M. M. Glazov, K. V. Kavokin, A. Lema{\^\i}tre, J. Bloch, M. Vladimirova, and D. Scalbert, Nondestructive measurement of nuclear magnetization by off-resonant faraday rotation, Phys. Rev. Lett. \textbf{111}, 087603 (2013).
    \bibitem{PhysRevB.15.5780} D. Paget, G. Lampel, B. Sapoval, and V. I. Safarov, Low field electron-nuclear spin coupling in gallium arsenide under optical pumping conditions, Phys. Rev. B \textbf{15}, 5780 (1977).
    \bibitem{d1974optical} M. I. Dyakonov and V. I. Perel, Optical orientation in a system of electrons and lattice nuclei in semiconductors. Theory, Sov. Phys. JETP \textbf{38}, 177 (1974).
    
   	\bibitem{note1} It has been pointed out that the Korringa mechanism also works for NSP relaxation in metallic samples with high doping concentration~\cite{ryzhov2015measurements}. However, this relaxation mechanism is not incorporated into our model, as our samples are considered insulating.
   	
   	    \bibitem{ryzhov2015measurements} I. Ryzhov, S. Poltavtsev, K. Kavokin, M. Glazov, G. Kozlov, M. Vladimirova, D. Scalbert, S. Cronenberger, A. Kavokin, A. Lema{\^\i}tre, et al., Measurements of nuclear spin dynamics by spin-noise spectroscopy, Appl. Phys. Lett. \textbf{106} (2015).
  
   	 \bibitem{kotur2016nuclear} M. Kotur, R. I. Dzhioev, M. Vladimirova, B. Jouault, V. L. Korenev, and K. V. Kavokin, Nuclear spin warm up in bulk $\mathit{n}$-GaAs, Phys. Rev. B \textbf{94}, 081201 (2016).
   	    
    \bibitem{rittmann2022unveiling}C. Rittmann, M. Yu. Petrov, A. N. Kamenskii, K. V. Kavokin, A. Yu. Kuntsevich, Yu. P. Efimov, S. A. Eliseev, M. Bayer, and A. Greilich, Unveiling the electron-nuclear spin dynamics in an $\mathit{n}$-doped InGaAs epilayer by spin noise spectroscopy, Phys. Rev. B \textbf{106}, 035202 (2022).

    \bibitem{millington2023nuclear} P. Millington-Hotze, S. Manna, S. F. Covre da Silva, A. Rastelli, and E. A. Chekhovich, Nuclear spin diffusion in the central spin system of a GaAs/AlGaAs quantum dot, Nat. Commun. \textbf{14}, 1 (2023).

    \bibitem{maletinsky2007dynamics} P. Maletinsky, A. Badolato, and A. Imamoglu, Dynamics of quantum dot nuclear spin polarization controlled by a single electron, Phys. Rev. Lett. \textbf{99}, 056804 (2007).

    \bibitem{adachi2012nuclear} S. Adachi, R. Kaji, S. Furukawa, Y. Yokoyama, and S. Muto, Nuclear spin depolarization via slow spin diffusion in single InAlAs quantum dots observed by using erase-pump-probe technique, J. Appl. Phys. \textbf{111}, 103531 (2012).

    \bibitem{maletinsky2009breakdown} P. Maletinsky, M. Kroner, and A. Imamoglu, Breakdown of the nuclear-spin-temperature approach in quantum-dot demagnetization experiments, Nat. Phys. \textbf{5}, 407 (2009).

    \bibitem{de1958relaxation} P-G. de Gennes, Sur la relaxation nucleaire dans les cristaux ioniques, J. Phys. Chem. Solids \textbf{7}, 345 (1958).

    \bibitem{PhysRevB.25.4444} D. Paget, Optical detection of NMR in high-purity GaAs: Direct study of the relaxation of nuclei close to shallow donors, Phys. Rev. B \textbf{25}, 4444 (1982).


    \bibitem{Shchepetilnikov_2013} A. V. Shchepetilnikov, Y. A. Nefyodov, I. V. Kukushkin, and W. Dietsche, Electron g-factor in GaAs/AlGaAs quantum wells of different width and barrier Al concentrations, J. Phys.: Conf. Ser. \textbf{456}, 012035 (2013).


    \bibitem{PhysRevB.90.205307} L.-P. Yan, M. Kurosawa, R. Kaji, G. Karczewski, S. Takeyama, and S. Adachi, Negative initial phase shift of Kerr rotation generated from the building-up process of resident electron spin polarization in a CdTe single quantum well, Phys. Rev. B \textbf{90}, 205307 (2014).


\end{thebibliography}

\end{document}